\documentclass[fleqn,usenatbib]{mnras}

\usepackage{newtxtext}

\usepackage[T1]{fontenc}
\usepackage{ae,aecompl}

\usepackage{graphicx}	
\usepackage{amsmath}	
\usepackage{amssymb}	
\usepackage[british]{babel} 
\usepackage{physics}
\usepackage{academicons}
\usepackage[dvipsnames]{xcolor}
\definecolor{orcidlogocol}{HTML}{A6CE39}
\newcommand{\orcid}[1]{\href{https://orcid.org/#1}{\textcolor[HTML]{A6CE39}{\aiOrcid}}}
\usepackage{booktabs}
\usepackage{makecell}
\usepackage{ulem}
\usepackage{extarrows}
\usepackage[nolist]{acronym}

\hypersetup{urlcolor=violet}

\newacro{AMR}{adaptive mesh refinement}
\newacro{BDM}{bound density maximum}
\newacro{CDM}{cold dark matter}
\newacro{cHMF}{cumulative halo mass function}
\newacro{CIC}{cloud-in-cell}
\newacro{CLPT}{convolutional Lagrangian perturbation theory}
\newacro{CMB}{cosmic microwave background}
\newacro{DGP}{Dvali-Gabadadze-Porrati}
\newacro{dHMF}{differential halo mass function}
\newacro{EFT}{effective field theory}
\newacro{FFT}{fast Fourier transformation}
\newacro{FLRW}{Friedmann-Lemaitre-Robertson-Walker}
\newacro{GELU}{Gaussian error linear unit}
\newacro{GR}{general relativity}
\newacro{GSM}{Gaussian streaming model}
\newacro{HAM}{halo abundance matching}
\newacro{HMF}{halo mass function}
\newacro{HOD}{halo occupation distribution}
\newacro{LPT}{Lagrangian perturbation theory}
\newacro{LRG}{luminous red galaxies}
\newacro{LSS}{large-scale structure}
\newacro{MCMC}{Markov chain Monte Carlo}
\newacro{MG}{modified gravity}
\newacro{NFW}{Navarro-Frenk-White}
\newacro{PDE}{partial differential equation}
\newacro{PDF}{probability distribution function}
\newacro{PM}{particle mesh}
\newacro{ReLU}{Rectified Linear Unit}
\newacro{RSD}{redshift space distortion}
\newacro{SHAM}{sub-halo abundance matching}
\newacro{SM}{streaming model}
\newacro{ST}{skewed student-T}
\newacro{STSM}{skew-T streaming model}
\newacro{TPCF}{two-point correlation function}

\newcommand{\rmm}{{\text{m}}}
\newcommand{\rmh}{{\text{h}}}
\newcommand{\rmhm}{{\text{hm}}}

\newcommand{\bdx}{{\boldsymbol{x}}}
\newcommand{\bdr}{{\boldsymbol{r}}}

\newcommand{\rmg}{{\text{g}}}
\newcommand{\rmgg}{{\text{gg}}}
\newcommand{\mathrms}{{\text{s}}}
\newcommand{\mathrmc}{{\text{c}}}

\definecolor{lightseagreen}{rgb}{0.13, 0.7, 0.67}

\newcommand{\dndm}{{\frac{\dd{n}}{\dd{m}}}}

\title[MG Emulator Halo Model I]{An emulator-based halo model in modified gravity -- I. The halo concentration-mass relation and density profile}

\author[C. Ruan et al.]{
Cheng-Zong Ruan$^{1,2}$, 
Carolina Cuesta-Lazaro$^{1,3}$, 
Alexander Eggemeier$^{4}$\thanks{Argelander Fellow},
Baojiu Li$^{1}\thanks{E-mail: baojiu.li@durham.ac.uk}$,
\newauthor
Carlton M. Baugh$^{1,3}$,
Christian Arnold$^{1}$,
Sownak Bose$^{1}$, 
C\'{e}sar Hern\'{a}ndez-Aguayo$^{5,6}$,
\newauthor
Pauline Zarrouk$^{7}$ and Christopher T. Davies$^{8}$ \\
$^{1}$Institute for Computational Cosmology, Department of Physics, Durham University, South Road, Durham DH1 3LE, UK\\
$^{2}$Institute of Theoretical Astrophysics, University of Oslo, 0315 Oslo, Norway\\
$^{3}$Institute for Data Science, Durham University, South Road, Durham DH1 3LE, UK\\
$^{4}$Argelander-Institut für Astronomie, Auf dem Hügel 71, D-53121 Bonn, Germany\\
$^{5}$Max-Planck-Institut fur Astrophysik, Karl-Schwarzschild-Str 1, D-85748 Garching, Germany\\
$^{6}$Excellence Cluster ORIGINS, Boltzmannstrasse 2, D-85748 Garching, Germany\\
$^{7}$Sorbonne Université, Université Paris Diderot, Sorbonne Paris Cité, CNRS/IN2P3, Laboratoire de Physique Nucléaire et de Hautes Energies (LPNHE), \\
\phantom{$^{7}$}4 place Jussieu, F-75252, Paris Cedex 5, France \\
$^{8}$Faculty of Physics, Ludwig-Maximilians-Universität, Scheinerstr. 1, 81679 Munich, Germany\\
}

\date{Accepted XXX. Received YYY; in original form \today}

\pubyear{2023}

\begin{document}
\label{firstpage}
\pagerange{\pageref{firstpage}--\pageref{lastpage}}
\maketitle

\begin{abstract}
In this series of papers we present an emulator-based halo model for the non-linear clustering of galaxies in modified gravity cosmologies.
In the first paper, we present emulators for the following halo properties: the halo mass function, concentration-mass relation and halo-matter cross-correlation function.
The emulators are trained on data extracted from the \textsc{FORGE} and \textsc{BRIDGE} suites of  $N$-body simulations, respectively for two modified gravity (MG) theories: $f(R)$ gravity and the DGP model, varying three standard cosmological parameters $\Omega_{\mathrm{m0}}, H_0, \sigma_8$, and one MG parameter, either $\bar{f}_{R0}$ or $r_{\mathrm{c}}$.
Our halo property emulators achieve an accuracy of $\lesssim 1\%$ on independent test data sets.
We demonstrate that the emulators can be combined with a galaxy-halo connection prescription to accurately predict the galaxy-galaxy and galaxy-matter correlation functions using the halo model framework.
\end{abstract}

\begin{keywords}
    dark energy -- large-scale structure of Universe -- cosmology: miscellaneous -- cosmology: theory.
\end{keywords}



\section{Introduction}
\label{sec:intro}

Ongoing and upcoming galaxy surveys, such as those that will be made with the  Dark Energy Spectroscopic Instrument \citep[\href{https://www.desi.lbl.gov/}{DESI};][]{DESI:2016arXiv161100036D}, \href{https://www.lsst.org/}the {Vera Rubin Observatory} \citep{LSST:2009arXiv0912.0201L} and \href{https://www.euclid-ec.org/}{\textit{Euclid}} \citep{Euclid:2011arXiv1110.3193L,AmendolaEuclid:2013LRR....16....6A,Troja:2022arXiv221109668T} will map the large scale structure (LSS) of the Universe with unprecedented statistical precision. 
Measurements of the large-scale structure can potentially be used to unveil the nature of the dark matter and dark energy, and to look for any deviation from the predictions of general relativity (GR). 
Theories of gravity beyond GR -- modified gravity (MG) models -- can  explain the observed accelerated expansion of the Universe without invoking a cosmological constant  \citep[e.g.][]{Koyama:2018IJMPD..2748001K,Ferreira:2019xrr}.
Studies of such models will not only shed light on the nature of the cosmic acceleration, but also serve as useful tests of GR on cosmic scales.


    The impact of modifications to GR has been well studied in terms of the cosmic expansion history and the large scale structure, i.e., at the background and linear perturbation levels. 
    In the late universe, the growth of LSS eventually enters the non-linear regime over a wide range of length scales, and linear theory predictions cease to be valid. 
    This point becomes even more acute in the context of MG, given that such models have intrinsically non-linear features, such as screening mechanisms and non-linear field equations for new degrees of freedom, which cannot be captured by linear theories \citep[e.g.][]{Li:2013MNRAS.428..743L}.
    Often a choice is made to exclude small scale data, thereby losing a wealth of information from high signal-to-noise ratio measurements.
    Such nonlinearities must be properly incorporated into theoretical modelling if one wishes to make the best use of the current and next generation cosmological surveys to constrain cosmological parameters and test gravity theories. 


    A fully non-linear treatment -- $N$-body simulations -- is essential to accurately solve the non-linear dynamics of cosmic structure formation \citep[see e.g.][for recent reivews]{Kuhlen:2012PDU.....1...50K,Angulo:2022LRCA....8....1A}. 
    The main hurdle of $N$-body simulations is their expensive computational cost. 
    A Monte Carlo Markov chain (MCMC) analysis, usually used to confront theoretical predictions with data, requires sampling at least $10^4$-$10^5$ models in the cosmological parameter space. 
    Probing such a large number of models using simulations is computationally prohibitive. 
    The situation is even worse for MG models, which usually involve partial differential equations governing the new physics.
    Current MG simulations can take between $2$ to $\mathcal{O}(10)$ times longer than standard $\Lambda$CDM simulations with the same specifications \citep[e.g.][]{Li:2011_ECOSMOG_code_paper,Arnold:2019NatAs...3..945A}.


    There are several approaches to dealing with the non-linear regime in addition to simulations.
    $N$-body simulation results can be used to develop phenomenological or semi-analytical fitting formulae to describe the  statistical properties of matter and dark matter haloes, such as the halo mass function calibrated by \citet{Tinker:2008ApJ...688..709T.HMF}, and the \textsc{halofit} prescription for the matter power spectrum \citep{Smith:2003MNRAS.341.1311S.halofit,Takahashi:2012ApJ...761..152T.halofit,Smith:2019MNRAS.486.1448S.halofit,Mead:2021MNRAS.502.1401M.HMCODE-2020} and bispectrum \citep{Takahashi:2020ApJ...895..113T.halofit}. 
    The most up-to-date version of \textsc{halofit} implemented by \citet{Mead:2021MNRAS.502.1401M.HMCODE-2020} achieves an accuracy of $5\%$ down to deeply non-linear scales.
    However, such parametric fits may no longer be fit for purpose with the advent of next-generation surveys that promise to reach measurements with per cent level precision.

    
    The halo model \citep{Neyman:1952ApJ...116..144N,Ma:2000ApJ...543..503M.halomodel,Peacock:2000MNRAS.318.1144P,Seljak:2000MNRAS.318..203S.HaloModel,Cooray:2002PhR...372....1C,Schmidt:2016PhRvD..93f3512S.halomodel,Philcox:2020PhRvD.101l3520P} is a successful analytical description of the LSS in the non-linear regime.
    In this framework, all matter, including galaxies and any other tracers, is assumed to reside within haloes.
    Then, the problem of predicting the clustering can be split into the following steps:
    \begin{itemize}
        \item the abundance of haloes as a function of halo mass, i.e. the halo mass function (HMF);
        \item the distribution of tracers around the halo centre, i.e. the halo density profile, usually assumed to be a Navarro-Frenk-White (NFW) profile \citep{Navarro:1996ApJ...462..563N.NFW,Navarro:1997ApJ...490..493N.NFW} specified by a halo mass-concentration relation; and
        \item the clustering of the haloes themselves, e.g., the halo two-point correlation function (TPCF) in configuration space and the halo auto power spectrum in Fourier space.
    \end{itemize}
    These basic properties of dark matter haloes constitute the halo model ingredients.
    The halo model provides a physically motivated description of the clustering statistics and is flexible enough to be extended to incorporate new physics such as massive neutrinos and baryonic feedback \citep[e.g.][]{Massara:2014JCAP...12..053M,Bose:2021MNRAS.508.2479B,Carrilho:2022MNRAS.512.3691C}, as well as models beyond $\Lambda$CDM and GR \citep[e.g.][]{Barreira:2014JCAP...04..029B,Lombriser:2014JCAP...03..021L,Hu:2018MNRAS.476L..65H.cham,Cataneo:2019MNRAS.488.2121C}.

    The halo model ingredients can be derived from analytic methods.
    For example, the HMF can be predicted using the spherical collapse model of the linear matter density field \citep{Press:1974ApJ...187..425P} and the excursion set formalism \citep{Bond:1991ApJ...379..440B,Sheth:2001MNRAS.323....1S}, whereby the dependence of the HMF on redshift and cosmology is expressed in terms of the root-mean-square fluctuations in the linear matter power spectrum. 
    \citet{Jenkins:2001MNRAS.321..372J} found that this universality of the HMF holds at an approximate level.
    As simulation predictions improved further, it was discovered that the redshift evolution of the mass function, even for $\Lambda$CDM, deviates from the universal prediction at the $5\text{-}10$ per cent level, and several new fitting formulae were proposed \citep[e.g.][]{Tinker:2008ApJ...688..709T.HMF,Courtin:2011MNRAS.410.1911C.HMF}.
    Moreover, the universality of the HMF is broken further in extensions of $\Lambda$CDM \citep[e.g.][for $w$CDM]{Bhattacharya:2011ApJ...732..122B.HMF} and modified gravity models \citep[e.g.][]{Schmidt:2010PhRvD..81f3005S,Lam:2012MNRAS.426.3260L,Li:2012MNRAS.421.1431L.baojiu,Lombriser:2013PhRvD..87l3511L,Cataneo:2016JCAP...12..024C,Gupta:2022PhRvD.105d3538G}.
    In order to obtain even tighter constraints on cosmological parameters and to test gravity theories, one therefore needs to proceed beyond the traditional approaches, given their limitations in accuracy and coverage of parameter space.


    In this series of papers, we develop simulation-based theoretical templates called \textit{emulators}, to obtain accurate predictions for basic halo properties as a function of halo mass and (modified gravity) cosmology, and to construct accurate predictions for clustering observables in preparation for ongoing and future galaxy surveys.
    There have been several previous works on the emulation of cosmological quantities in the $\Lambda$CDM model and its extensions, such as \citet{Heitmann:2006ApJ...646L...1H.emu,Habib:2007PhRvD..76h3503H}, the Coyote Universe \citep{Heitmann:2010ApJ...715..104H,Heitmann:2009ApJ...705..156H,Lawrence:2010ApJ...713.1322L}, PkANN \citep{Agarwal:2012MNRAS.424.1409A,Agarwal:2014MNRAS.439.2102A}, the Mira-Titan Universe \citep{Heitmann:2016ApJ...820..108H.Mira-Titan1,Lawrence:2017ApJ...847...50L.Mira-Titan2,Bocquet:2020ApJ...901....5B.MTHMF}, \citet{Kwan:2013ApJ...768..123K,Kwan:2015ApJ...810...35K}, \href{https://aemulusproject.github.io/index.html}{\textsc{Aemulus}} \citep{DeRose:2019ApJ...875...69D.AemulusI,McClintock:2019ApJ...872...53M.AemulusHMF,Zhai:2019ApJ...874...95Z}, \href{https://darkquestcosmology.github.io/}{\textsc{Dark Quest}} \citep{Nishimichi:2019ApJ...884...29N,Kobayashi:2020PhRvD.102f3504K,Miyatake:2021arXiv210100113M.DQ,Cuesta-Lazaro:2022arXiv220805218C}, \href{https://github.com/JDonaldM/Matryoshka}{\texttt{matryoshka}} \citep{Donald-McCann:2022MNRAS.511.3768D} and \href{https://abacussummit.readthedocs.io/en/latest/}{\textsc{AbacusSummit}} \citep{Maksimova:2021MNRAS.508.4017M.ABACUSSUMMITOverview,Yuan:2022MNRAS.515..871Y}, as well as in non-standard cosmologies, such as \citet{Winther:2019PhRvD.100l3540W,Ramachandra:2021PhRvD.103l3525R,Arnold:2022MNRAS.515.4161A.FORGE_1,Brando:2022JCAP...09..051B,Harnois:2022arXiv221105779H.MGLenS}.

    To build emulators we use the machine-learning interpolation technique of neural networks, which allows us to predict halo properties for any given cosmology within the range of parameters covered by the training data set.
    We use the \textsc{FORGE} (F-Of-R Gravity Emulator) and \textsc{BRIDGE} (BRaneworld-Inspired D\textsc{gp} Gravity Emulator) modified gravity $N$-body simulations described in \citet{Arnold:2022MNRAS.515.4161A.FORGE_1}, which together cover a very broad range of parameters in two MG theories: $f(R)$ gravity and the DGP model.
    The emulated halo properties incorporate all the complicated effects on non-linear scales, such as the non-linear halo bias, the halo exclusion effect and the screening mechanism.

    Following the spirit of the \textsc{Dark Quest} project \citep{Nishimichi:2019ApJ...884...29N,Cuesta-Lazaro:2022arXiv220805218C}, we do not perform an end-to-end emulation of galaxy clustering statistics in the joint parameter space of cosmological and galaxy-halo connection models.
    Instead, we develop emulators for each halo property separately, and assemble these ingredients within the halo model framework to construct analytical predictions of galaxy clustering statistics.
    This emulator-based halo model gives us the flexibility to insert different prescriptions of galaxy-halo connection for different galaxy samples.
    Moreover, emulators for basic halo properties themselves are very useful.
    For example, calibrating the cosmology dependence of the HMF is crucial to control the systematic uncertainty in galaxy cluster abundance studies \citep[e.g.][]{McClintock:2019ApJ...872...53M.AemulusHMF,Bocquet:2020ApJ...901....5B.MTHMF}.


The layout of this paper is as follows. 
In Section~\ref{sec:mg_theories}, we present a short description of the modified gravity theories studied here and a brief overview of the \textsc{FORGE} and \textsc{BRIDGE} $N$-body simulation suites.
In Section~\ref{sec:dataset}, we outline the measurement and post-processing of the halo properties from the simulations.
Section~\ref{sec:nn} describes the construction of the halo property emulators using neural networks, and Section~\ref{sec:results} shows their performance in reproducing the simulation results.
In Section~\ref{sec:applications}, we demonstrate how these emulators can be combined with a galaxy-halo connection prescription to predict galaxy statistics.

Throughout this paper, we use $\log$ to denote the base-$10$ logarithm, $\log \equiv \log_{10}$, and $\ln$ to indicate the natural logarithm.
Unless otherwise stated, we use a subscript $0$ to denote the present-day value of a physical quantity and an overbar for the background value of a quantity.

\section{Modified Gravity Theories and Simulations}
\label{sec:mg_theories}

We briefly describe the two modified gravity models analysed in this work, $f(R)$ gravity \citep{Hu:2007PhRvD..76f4004H} and the \ac{DGP} brane-world models \citep{Dvali:2000PhLB..485..208D.DGP}. 
These are two of the most widely studied MG models and, as we discuss below, are representative examples of the two main classes of screening mechanisms, which make them good test-beds for generic MG models. 
For more detailed descriptions of these models, we refer the reader to \citet{Sotiriou:2010RvMP...82..451S} and \citet{DeFelice:2010LRR....13....3D} for $f(R)$ gravity, and \citet{2003JCAP...11..014S} and \citet{maartens2010brane} for DGP models.

\subsection[f(R) gravity]{$f(R)$ gravity}
\label{subsec:fRgrav}

The $f(R)$ gravity is a generalisation of Einstein's general relativity. In $f(R)$ gravity, the Einstein-Hilbert action in GR has an additional term, which is a function of the Ricci scalar $R$,
\begin{align}
    S = \int \dd^4 x \sqrt{-g} \, \qty{\frac{M^2_{\rm Pl}}{2} \qty[ R + f(R)] + \mathcal{L}_m } \ , \label{eqn:fRaction_idv}
\end{align}
where $M_{\rm Pl} = (8 \pi G)^{-1/2}$ is the reduced Planck mass, $G$ is Newton's constant, $g$ is the determinant of the metric $g_{\mu \nu}$ and $\mathcal{L}_m$ is the Lagrangian density for matter fields. 
Varying the action with respect to the metric $g_{\mu\nu}$ gives the modified Einstein equation,
\begin{align}
    G_{\mu \nu} + f_R R_{\mu \nu} - \qty(\frac{1}{2} f - \square f_R ) g_{\mu \nu} - \nabla_\mu \nabla_\nu f_R = 8\pi G T^{\rmm}_{\mu \nu} , \label{eqn:einstein_field_eqn_for_fR_efsd}
\end{align}
where 
\begin{equation}
G_{\mu\nu} \equiv R_{\mu\nu} - \frac{1}{2}g_{\mu\nu}R,
\end{equation} 
is the Einstein tensor, $f_R \equiv \dd f(R) / \dd R$, $\nabla_\mu$ is the covariant derivative corresponding to the metric $g_{\mu\nu}$, $\square \equiv \nabla^\alpha \nabla_\alpha$ and $T^{\rmm}_{\mu\nu}$ is the energy momentum tensor for matter.

Equation~\eqref{eqn:einstein_field_eqn_for_fR_efsd} is a fourth-order partial differential equation in $g_{\mu \nu}$. This equation can also be  considered as the standard Einstein equation in GR with a new dynamical degree of freedom, $f_R$, which is dubbed the \textit{scalaron}  \citep[e.g.,][]{2011PhRvD..83d4007Z}. 
The equation of motion of $f_R$ can be obtained by taking the trace of Equation~\eqref{eqn:einstein_field_eqn_for_fR_efsd}:
\begin{align}
    \square f_R = \frac{1}{3} \qty(R - f_R R + 2f + 8 \pi G \rho_{\mathrm{m}}) \ , \label{eqn:fR_EoM_ewfasd}
\end{align}
where $\rho_{\mathrm{m}}$ is the matter density.

For cosmological simulations in standard gravity, the Newtonian limit is commonly adopted. This includes the approximations that the  gravitational and scalar fields are weak (such that their higher-order terms can be neglected) and quasi-static (so that the time derivatives of the fields can be neglected compared to their spatial derivatives). Most modified gravity simulations (including the ones used in this work) adopt this assumption. In the context of $f(R)$ gravity and the Newtonian limit, the modified Einstein equation \eqref{eqn:einstein_field_eqn_for_fR_efsd} becomes
\begin{align}
    {\nabla}^2 \Phi &\approx \frac{16 \pi G}{3} a^2 (\rho_{\mathrm{m}} - \bar{\rho}_{\mathrm{m}}) + \frac{1}{6} a^2 \qty[R(f_R) - \bar{R}] \ , \\
\intertext{and the equation of motion of the scalaron reduces to}
    {\nabla}^2 f_R &\approx -\frac{1}{3} a^2 \qty[ R(f_R) - \bar{R} + 8 \pi G (\rho_{\mathrm{m}} - \bar{\rho}_{\mathrm{m}})] \ , \label{eqn:fR_scalar_field_equation}
\end{align}
where $\Phi$ is the Newtonian potential, ${\nabla}$ is the 3-dimensional gradient operator, and an overbar denotes the cosmic mean of a quantity.

An $f(R)$ gravity model is fully specified by the functional form of $f(R)$.
Here, we adopt the well-studied Hu-Sawicki model \citep{Hu:2007PhRvD..76f4004H}, which is given by
\begin{align}\label{eq:HS_fr}
    f(R) = -m^2 \frac{c_1 (-R/m^2)^n}{c_2 (-R/m^2)^n + 1} \ ,
\end{align}
where $m^2 \equiv \Omega_{\rmm0} H_0^2$, and $c_1, c_2$ and $n$ are free parameters. 
The parameter $n$ is a positive number, which is set to $n=1$ in this work as in most previous studies of this model \citep[however see e.g.,][for some examples of $n\neq1$]{Li:2011uw,Ramachandra:2021PhRvD.103l3525R,Ruan:2022JCAP...05..018R}. 
With this functional form, we have
\begin{align}\label{eq:generalised_HS_fr}
    f_R = -\left| \bar{f}_{R0} \right|\left(\frac{\bar{R}_0}{R}\right)^{n+1},
\end{align}
where $\bar{R}_0$ and $\bar{f}_{R0}$ are, respectively, the present-day values of the background Ricci scalar and $\bar{f}_R$. 
For brevity, we will adopt the following nomenclature to label models: the model with $-\log_{10}\left(|\bar{f}_{R0}|\right) = 5.5$ will be called F5.5, and so on.

The remaining free parameter of the theory is the background value of the scalar field $f_R$ at redshift $z = 0$, $\bar{f}_{R0}$. 
With a suitable choice of this parameter, $f(R)$ gravity reverts to GR in high-density regions -- this is necessary to be consistent with solar system tests through the associated chameleon mechanism \citep{Khoury:2004PhRvL..93q1104K,Khoury:2004PhRvD..69d4026K}.
A larger value of $|\bar{f}_{R0}|$ means a stronger deviation from standard gravity. 
The F5 model is in slight tension with small-scale tests, see, e.g., \citet{2014AnP...526..259L} for a recent review of current cosmological constraints on $\bar{f}_{R0}$. 
But since we aim to test gravity on cosmic scales, models with such strength of MG are nevertheless still valuable to study: given their stronger deviations from GR compared to models with smaller $|\bar{f}_{R0}|$, they can lead to important insights into how the deviations from GR can affect large-scale cosmological observables such as weak lensing and galaxy clustering statistics. 
In order to fully explore the gravity testing capacities of upcoming cosmological observations, it is important to gain a detailed understanding of how these measures are influenced by possible modifications to gravity.

\subsection{The Dvali-Gabadadze-Porrati (DGP) model}
\label{subsec:DGPmod}

In the Dvali-Gabadadze-Porrati braneworld model \citep{DGP:2000PhLB..485..208D}, the universe is a four-dimensional brane embedded in a five-dimensional space-time (called the bulk). 
The gravitational action in this model is given by 
\begin{align}
    S = \int_{\text{brane}} \dd[4]{x} \sqrt{-g} \qty(\frac{R}{16 \pi G}) + \int_{\text{bulk}} \dd[5]{x} \sqrt{-g^{(5)}} \qty(\frac{R^{(5)}}{16 \pi G^{(5)}}), \label{eqn:grav_act_DGP}
\end{align}
where a superscript $^{(5)}$ denotes the quantity in the five-dimensional bulk. 
This model has a self-accelerating branch of solution (sDGP), which gives a natural explanation for the cosmic acceleration. 
However, the sDGP branch suffers from the ghost problems \citep{2007CQGra..24R.231K} and cannot be considered as a physical model.
Moreover, its predictions have been found to be inconsistent with observations such as \ac{CMB} and local measurements of the Hubble parameter $H_0$ \citep[e.g.,][]{2007PhRvD..75f4003S,2008PhRvD..78j3509F}.

The so-called normal branch DGP (nDGP) gravity \citep{2007CQGra..24R.231K} cannot accelerate cosmic expansion by itself, so in order to explain the cosmological observations it has to introduce additional dark energy components.
This model is nevertheless still of interest as a useful toy model that features the Vainshtein screening mechanism \citep{Vainshtein:1972PhLB...39..393V}.
Here, we assume that there is an additional non-clustering dark energy component in this model, which results in its expansion history being identical to that of $\Lambda$CDM. 
The nDGP model provides an explanation of why gravity is much weaker than the other fundamental forces \citep{maartens2010brane}: all matter species are assumed to be confined to the brane, while gravity could propagate through (leak into) the extra spatial dimensions. 
There is one new free parameter in the nDGP model, which can be defined as the ratio of $G^{(5)}$ and $G$, and is known as the crossover scale, 
\begin{align}
    r_c \equiv \frac{1}{2} \frac{G^{(5)}}{G}\,.
\end{align}

The modified Friedmann equation in the normal branch DGP model is given by
\begin{align}
    \frac{H(a)}{H_0} = \sqrt{\Omega_{\rmm0} a^{-3} + \Omega_{\rm DE0}(a) + \Omega_{\rm rc}} - \sqrt{\Omega_{\rm rc}} \ , \label{eqn:Friedmann_eqn_nDGP}
\end{align}
where $\Omega_{\rm rc} \equiv 1 / (4H_0^2 r_c^2)$, and $\Omega_{\rm DE0}$ is the density parameter of the additional dark energy component.
The dimensionless quantity $H_0 r_c$ is used to quantify departures from the standard gravity.
If $H_0 r_c \to \infty$ then Eqn.~\eqref{eqn:Friedmann_eqn_nDGP} returns to the $\Lambda$CDM case.
A larger value of $H_0 r_c$ means a weaker deviation from GR.
Hereafter, an nDGP model with $H_0 r_c = X$ will be referred to as N$X$.
For example, a model with $H_0r_c=1$ is called N1.

The modified Poisson equation and the scalar field equation are given by \citep{2007CQGra..24R.231K},
\begin{equation}\label{eq:poisson_nDGP}
\nabla^2 \Phi = 4\pi G a^2 \delta\rho_{\rm m} + \frac{1}{2}\nabla^2\varphi\,,
\end{equation}
and
\begin{equation}\label{eq:phi_dgp}
\nabla^2 \varphi + \frac{r_c^2}{3\beta\,a^2c^2} \left[ (\nabla^2\varphi)^2
- (\nabla_i\nabla_j\varphi)^2 \right] = \frac{8\pi\,G\,a^2}{3\beta} \delta\rho_{\rm m}\,,
\end{equation}
where $\varphi$ is a new scalar degree of freedom, $\delta\rho_{\rm m} = \rho_{\rm m} - \bar{\rho}_{\rm m}$ and 
\begin{align}
\beta(a) &\equiv 1 + 2 H\, r_c \left ( 1 + \frac{\dot H}{3 H^2} \right ) \notag \\
&= 1 + \frac{\Omega_{\rmm0}a^{-3} + 2\Omega_\Lambda0}{2\sqrt{\Omega_{\rm rc}(\Omega_{\rmm0}a^{-3} + \Omega_{\Lambda0})}}. \label{eq:beta_dgp}
\end{align}

\subsection{Modified gravity $N$-body simulations}
\label{subsec:sims}

To construct emulators for dark matter halo properties, we use the \textsc{FORGE} and \textsc{BRIDGE} suites of $N$-body simulations \citep{Arnold:2022MNRAS.515.4161A.FORGE_1}, covering $49$ $f(R)$ gravity and $49$ DGP models, along with $49$ $\Lambda$CDM counterparts.
The simulations were performed using $1024^3$ dark matter particles in a cube of side  $500\,h^{-1}\,\mathrm{Mpc}$ (hereafter the high-resolution runs, denoted HR) or $1500\,h^{-1}\,\mathrm{Mpc}$ (low-resolution runs, labelled LR), using the modified gravity version of the \textsc{Arepo} cosmological simulation code \citep{Springel:2010_AREPO_code_paper,Arnold:2019_MGAREPO_code_paper,Weinberger:2020ApJS..248...32W.arepo}.
The mass resolutions of the HR and LR runs are $9.1 \times 10^{9}$ and $1.5 \times 10^{12} (\Omega_{\rmm0}/0.3)\,h^{-1}\,M_{\odot}$, respectively.
The gravitational softening lengths of simulations are $15$ (HR) and $75\,h^{-1}\mathrm{kpc}$ (LR).
The initial conditions were generated using second-order Lagrangian perturbation theory (2LPT, \citet{Crocce:2006MNRAS.373..369C.2LPTIC}) at $z_{\mathrm{init}} = 127$.
Each cosmology (also called ``node'') has two independent realisations with the pairs of initial conditions selected to minimise the sample variance on large scales over the realisations. 
All nodes were initialised with the same (two) random seeds.
See Section~3.2 of \citet{Arnold:2022MNRAS.515.4161A.FORGE_1} for a detailed description.

\begin{figure}
    \centering
    \includegraphics[width=\columnwidth]{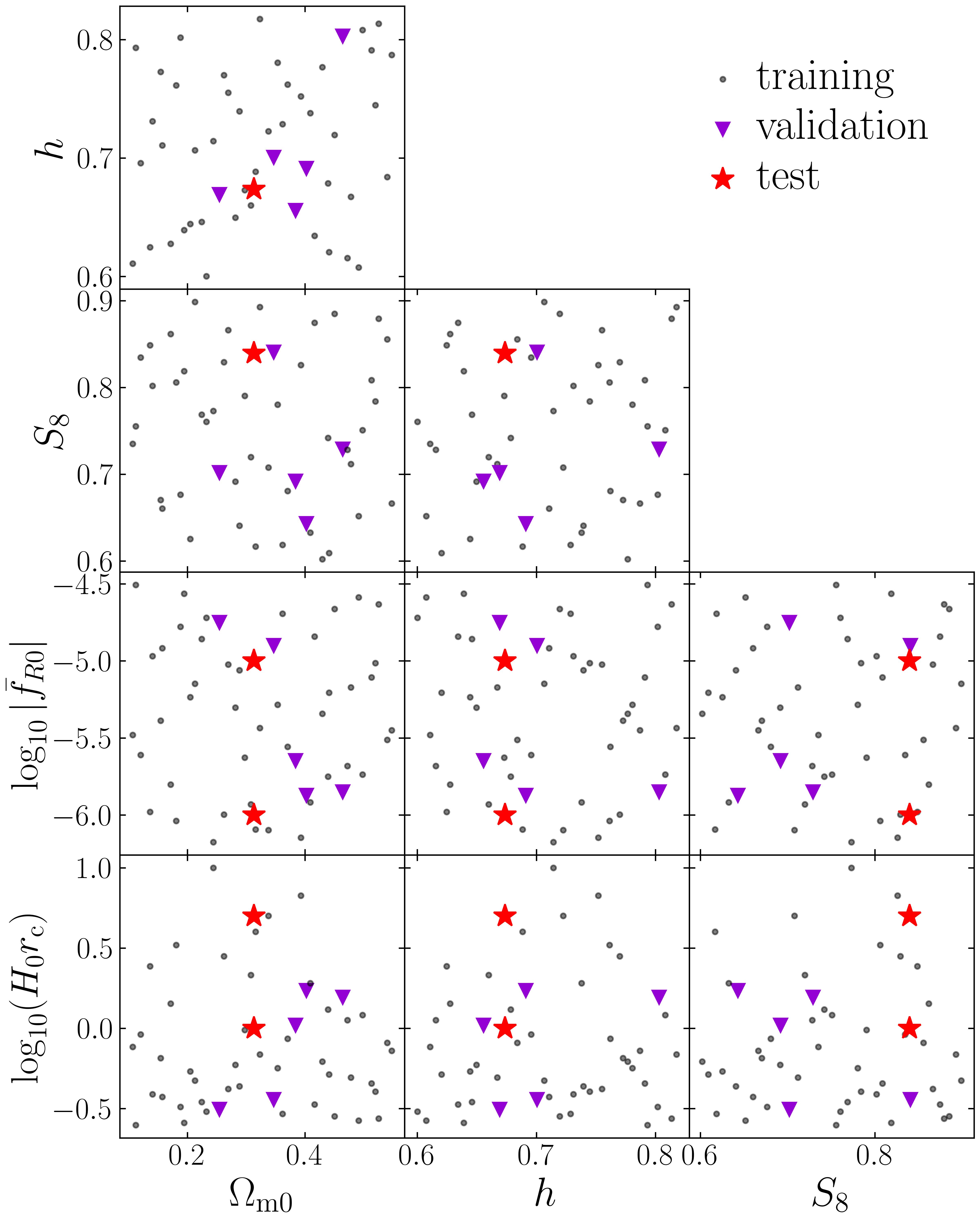}
    \caption{
    Visualisation of the cosmological parameters $\boldsymbol{\mathcal{C}}$ (Equations~\eqref{eqn:cos_LCDM}-\eqref{eqn:cos_DGP}) covered in the \textsc{FORGE} and \textsc{BRIDGE} simulations.  
    The $44$ training cosmologies, $5$ validation and $2$ test cosmologies are shown as  black dots, purple triangles  and red stars, respectively.
    $\Lambda$CDM models corresponding to  $\log{|\bar{f}_{R0}|} = -\infty$ and $\log{(H_0 r_{\mathrm{c}}) = \infty}$ are {\it not} shown in the last two rows.
    \href{https://github.com/chzruan/HM_emulator_paper1_figures/blob/main/params_FORGE_BRIDGE.py}{(source code)}
    }
    \label{fig:FORGEparamspace}
\end{figure}

The cosmological parameters were drawn from a Latin hypercube designed to efficiently sample a 4-dimensional parameter space (or a 3-dimensional space in the case of  $\Lambda$CDM), as shown in Fig.~\ref{fig:FORGEparamspace}, following a similar approach to that used in the cosmo-SLICS project \citep{Harnois-Deraps:2019A&A...631A.160H}.
Since \textsc{FORGE} and \textsc{BRIDGE} were partly designed to emulate weak lensing statistics, they sampled directly in the composite structure growth parameter
\begin{align}
    S_8 \equiv \sigma_8 \sqrt{\frac{\Omega_{\rmm0}}{0.3}},
\end{align}
instead of the physical matter fluctuation amplitude parameter $\sigma_8$.
The use of $S_8$ accounts better for the degeneracy between $\Omega_{\rmm0}$ and $\sigma_8$ in the cosmic shear analysis.
The $49$ nodes form the training data set for each gravity model.
It is common practice to use a small portion (called validation set) of the training set to determine whether the process has finished.
We choose the nodes 11, 13, 22, 34, 36 as the validation set.

The following three standard cosmological parameters and one MG parameter are varied,
\begin{align}
    \boldsymbol{\mathcal{C}} &= \qty{\Omega_{\mathrm{m0}}, h, S_8}, \ &&\text{for $\Lambda$CDM}, \label{eqn:cos_LCDM}\\
    \boldsymbol{\mathcal{C}} &= \qty{\Omega_{\mathrm{m0}}, h, S_8, \log_{10}|\bar{f}_{R0}|},\ &&\text{for $f(R)$}, \label{eqn:cos_fR} \\
    \boldsymbol{\mathcal{C}} &= \qty{\Omega_{\mathrm{m0}}, h, S_8, \log_{10}(H_0 r_{\mathrm{c}})},\ &&\text{for DGP},  \label{eqn:cos_DGP}
\end{align}
where $h \equiv H_0 / (100\,\mathrm{km}\,\mathrm{s}^{-1}\,\mathrm{Mpc}^{-1})$ and $H_0$ is the present day Hubble constant.
The range of the parameters explored is 
\begin{equation}
\begin{split}
    0.11 &< \Omega_{\rmm0} < 0.54 \\
    0.61 &< h < 0.81 \\
    0.6  &< S_8 < 0.9 \\
    -6.2 &< \log_{10}|\bar{f}_{R0}| < -4.6 \\
    -0.45 &< \log_{10}(H_0 r_{\mathrm{c}}) < 1.0.
\end{split} \label{eqn:param_range_slhc}
\end{equation}
The density parameter of baryons was fixed to $\Omega_{\mathrm{b0}} = 0.049199$ and massive neutrinos are ignored.
The dark energy density is given by
\begin{align}
    \Omega_{\Lambda0} = 1 - \Omega_{\rmm0}.
\end{align}
The remaining cosmological parameter is the slope of the primordial curvature power spectrum normalised at $0.05\,\mathrm{Mpc}^{-1}$, which is $n_{\mathrms} = 0.9652$ \citep{Arnold:2022MNRAS.515.4161A.FORGE_1}.

We also need test data set to assess the performance of the emulators independently.
In both cases, the test set consists of two models.
The MG test cases are F5 and F6 for $f(R)$ gravity, and for DGP they are N1 and N5, and they share the same cosmological parameters, given by the fiducial \textit{Planck} cosmology \citep[][]{Planck2018}, 
\begin{align}
    &\Omega_{\mathrm{m0}} = 0.31315,\quad \Omega_{\Lambda0} = 0.68685,\quad \Omega_{\mathrm{b0}} = 0.049199, \notag \\
    &h = 0.6737,\quad \sigma_8 = 0.82172,\quad n_{\mathrms} = 0.9652.  \label{eqn:fig_planck_cos_forge}
\end{align}
Each test model has $8$ realisations.

The dark matter halo catalogues were obtained using the \textsc{Subfind} halo finder  \citep{Springel2001}.
The haloes were first identified using a fast parallel friends-of-friends (FOF) algorithm with link length set to $b = 0.168$ times the mean interparticle separation. 
Spherical overdensity halo catalogues are then built out from the potential minimum of each FOF halo. 
The halo mass definition adopted is
\[
    M_{\mathrm{200c}} \equiv \frac{4\pi}{3} (R_{\mathrm{200c}})^3 \times 200 \rho_{\mathrm{crit}},
\]
where $\rho_{\mathrm{crit}}(z) \equiv 3H^2(z) / (8\pi G)$ is the critical density of the Universe, and $R_{\mathrm{200c}}$ is the spherical halo radius within which the spherically averaged mass density equals $200\rho_{\mathrm{crit}}$.
Only main haloes with masses above $10^{12}\,h^{-1} M_{\odot}$ from the HR simulations are considered in this work.
The halo catalogues at 
\begin{align*}
    z = 0.00, 0.25, 0.50, 0.75, 1.00, 1.25, 1.50, 1.75 \ \text{and}\ 2.00
\end{align*} 
are available for all nodes.
Besides these common redshifts, Arnold et~al. saved particle snapshots and halo catalogues at pre-selected redshifts to construct past light-cones for weak-lensing analysis, therefore enabling the emulation of halo properties as a function of redshift.
In the main text, we present the performance of the emulators at redshift zero, since the measurement and emulation at other redshifts are performed in the same way.

\section{Data Set}
\label{sec:dataset}
In this section we describe the measurement and post-processing of the halo properties from the simulations, including the halo mass function, concentration-mass relation and halo-matter cross-correlation function.

\subsection{Halo Mass Functions}
\label{subsec:hmf_data}

The differential halo mass function (dHMF) quantifies the number density of haloes as a function of halo mass for a given cosmology $\mathcal{C}$ and redshift.
It is denoted as
\begin{align}
    \dv{n(M; z, \boldsymbol{\mathcal{C}})}{M} \quad \text{or} \quad \dv{n(\log M; z, \boldsymbol{\mathcal{C}})}{\log M}.
\end{align}
In the cumulative form, the cumulative HMF (cHMF) gives the number density of haloes above a given mass threshold $M$,
\begin{align}
    n(>M; z, \boldsymbol{\mathcal{C}}) = \int_{M}^{\infty} \dd{m} \dv{n(m; z, \boldsymbol{\mathcal{C}})}{m}. 
\end{align}

The HMF is measured by creating a histogram of the halo mass, which is affected by shot noise and sample variance, especially for massive haloes.
Also, HMFs span many orders of magnitude in abundance, typically from $10^{-3}$ to $10^{-8}\,(h^{-1}\mathrm{Mpc})^{-3}$.
Taking the logarithm of the HMF to reduce the dynamic range does not help much, since the interpolation errors in the logarithmic quantity would be exponentially amplified.
To overcome these problems, a commonly used approach (e.g. followed by   \citealt{McClintock:2019ApJ...872...53M.AemulusHMF,Nishimichi:2019ApJ...884...29N,Cuesta-Lazaro:2022arXiv220805218C}) is to fit the measured HMF using fitting formulae like those proposed by  \citet{Jenkins:2001MNRAS.321..372J,Tinker:2008ApJ...688..709T.HMF},  and then to emulate the mass and cosmology dependence of the fitting parameters.
However, the performance of such fitting functions in MG simulations is not guaranteed \citep[e.g.][]{Schmidt:2009PhRvD..79h3518S,Gupta:2022PhRvD.105d3538G} and therefore may cause systematic errors.

We adopt an alternative method to emulate the HMF.
The main ideas include: 
(1) Emulating the ratio between the simulation result for the HMF and a realistic fitting formula to reduce the dynamic range.
(2) Considering the cumulative HMF instead of the differential one to allow for smaller steps in halo mass, thereby providing more training data.

\begin{figure}
    \centering
    \includegraphics[width=0.95\columnwidth]{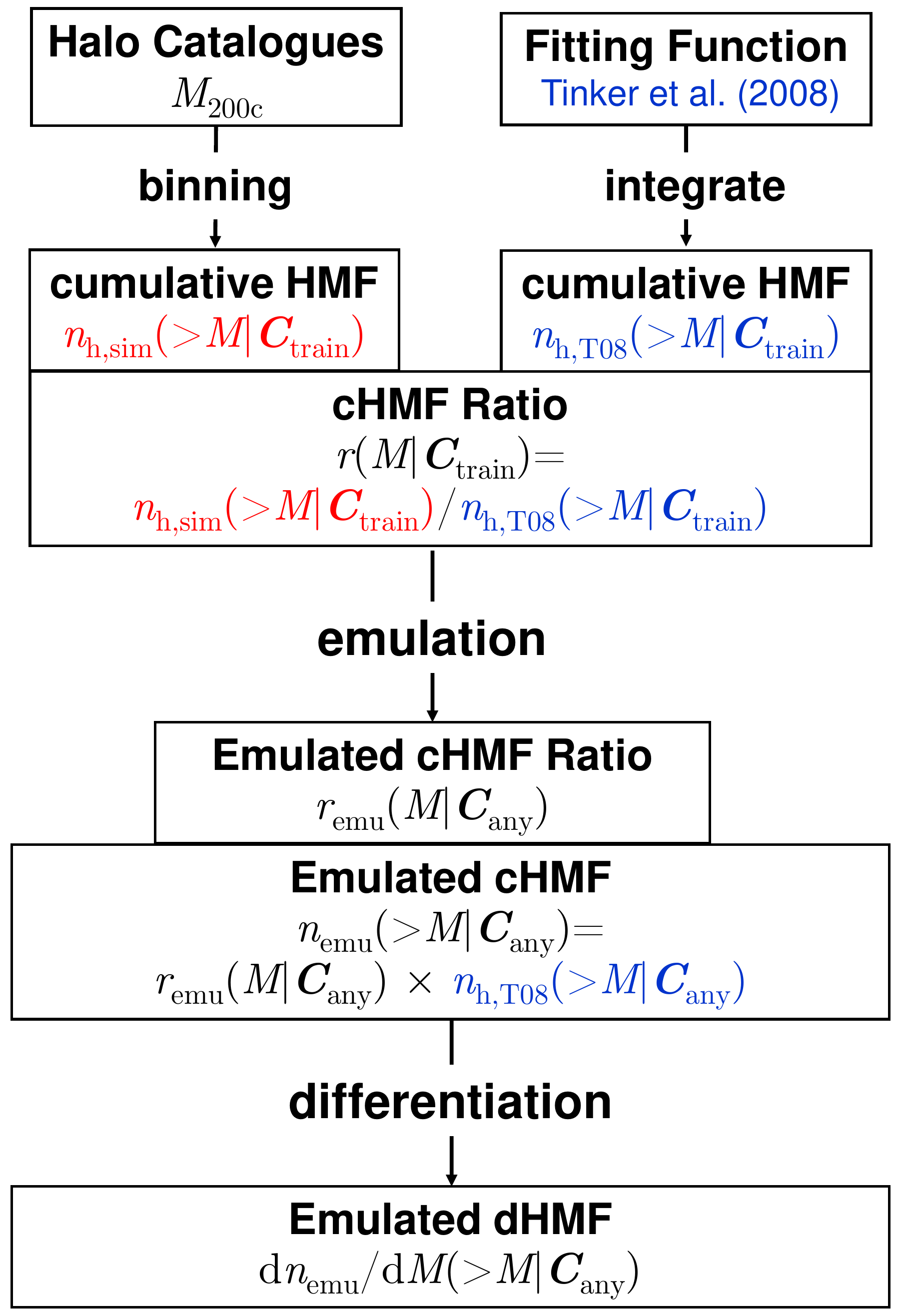}
    \caption{
        Flow chart illustrating the process of emulating halo mass functions.
        We start from measuring the cumulative HMF from the simulations in the training data set.  
        For each simulation, we calculate the cHMF predicted by the fitting formula of \citet{Tinker:2008ApJ...688..709T.HMF} for the same cosmology $\boldsymbol{\mathcal{C}}$.
        We then interpolate the ratios between the two cHMFs across parameter space using neural networks.
        To obtain the commonly used differential HMF for any given cosmology $\boldsymbol{\mathcal{C}}_{\mathrm{any}}$, we can calculate the ratio, multiply the ratio by the \citet{Tinker:2008ApJ...688..709T.HMF} function and take the derivative.
    }
    \label{fig:HMF_emu_pipeline}
\end{figure}

\citet{Tinker:2008ApJ...688..709T.HMF} (hereafter T08) derived fitting functions for the HMF applicable over a wide range of halo masses and halo definitions, with a precision of $\lesssim5\%$.
We work with the ratio of the cumulative HMFs between the simulation measurements and the T08 fitting formulae\footnote{Numerically implemented in the Python module \href{https://hmf.readthedocs.io/en/latest/index.html}{\texttt{hmf}}  \citep{Murray:2013A&C.....3...23M.HMFcalc,Murray2021A&C....3600487M.THEHALOMOD}.},
\begin{align}
    r(M; \boldsymbol{\mathcal{C}}, z) \equiv \frac{n_{\rmh}^{\mathrm{sim}}(>M; \boldsymbol{\mathcal{C}}, z)}{n_{\rmh}^{\mathrm{T08}}(>M; \boldsymbol{\mathcal{C}}, z)}. \label{eqn:hmf_ratio}
\end{align}
The HMF ratios are then interpolated across parameter space using neural networks to construct the emulator.
To obtain the differential HMF for any given cosmology $\boldsymbol{\mathcal{C}}_{\text{any}}$, one can calculate the ratio for this cosmology using the trained emulator, multiply it with the T08 HMF and take the  derivative.
The emulation process is sketched in Fig.~\ref{fig:HMF_emu_pipeline}.

\begin{figure}
    \centering
    \includegraphics[width=0.92\columnwidth]{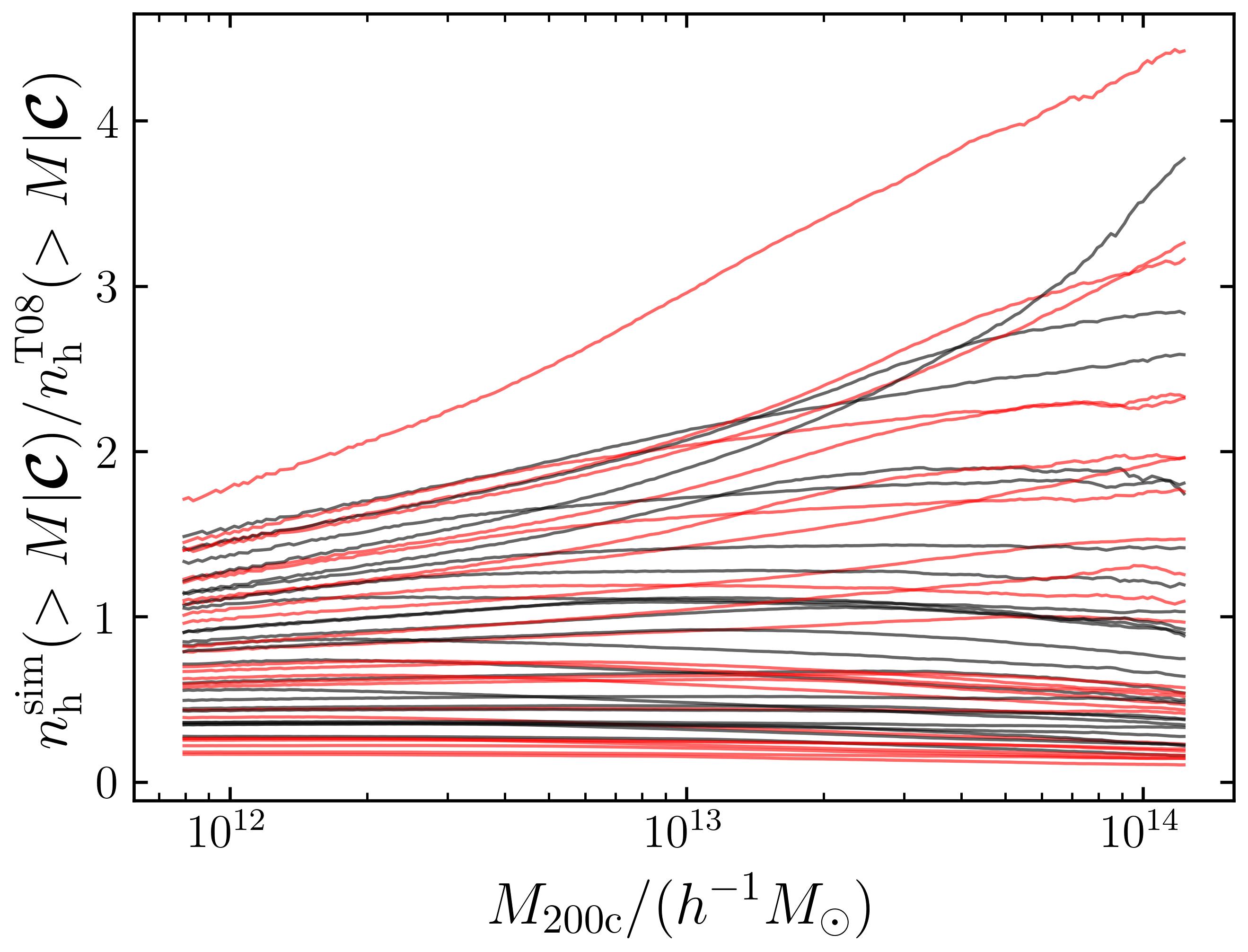}
    \caption{Cumulative halo mass function (cHMF) ratios between simulation measurements for $49$ $f(R)$ gravity models and the fitting formula calibrated by \citet{Tinker:2008ApJ...688..709T.HMF} for the same cosmology (except for the MG parameter $\bar{f}_{R0}$, which is set to zero), at redshift $0$. 
    The dynamic range of the ratios are significantly decreased compared with cHMFs themselves, therefore increasing the emulation accuracy.
    (\href{https://github.com/chzruan/HM_emulator_paper1_figures/blob/main/HMF_ratio.py}{source code})
    }
    \label{fig:cHMF_ratio}
\end{figure}

In each snapshot, we measure the number densities of haloes more massive than a series of masses, beginning at $10^{12}\,h^{-1}M_{\odot}$ and increasing in steps with a bin width of $\Delta\log{[M / (h^{-1}M_{\odot})]} = 0.02$.
The maximum mass varies across redshifts.
Fig.~\ref{fig:cHMF_ratio} presents the ratios of HMFs for $49$ $f(R)$ gravity simulations at $z = 0$.
The ratios are gently varying functions over a lower dynamic range than HMFs themselves, therefore resulting in a higher emulation accuracy.

\subsection{The individual halo density profile and the concentration-mass relation}
\label{subsec:halo_profile_individual_nfw}

One of the most remarkable discoveries from cosmological $N$-body simulations was that  dark matter haloes display a universal density profile \citep{Navarro:1996ApJ...462..563N.NFW,Navarro:1997ApJ...490..493N.NFW,Wang:2020Natur.585...39W}, from the host haloes of dwarf galaxies to those of massive galaxy clusters.
Specifically, it was shown that the spherically averaged density profile of individual relaxed haloes can be described by the well-known Navarro, Frenk \& White (NFW) profile. 
The NFW profile is described by two parameters, the characteristic density and scale radius of a halo, or equivalently the halo mass and  concentration,
\begin{align}
    \rho_{\mathrm{NFW}} (r | r_{-2}, \rho_{-2}) = \frac{ 4\rho_{-2} }{\displaystyle \qty(\frac{r}{r_{-2}}) \qty(1 + \frac{r}{r_{-2}})^2 }, \label{eqn:NFW_profile_exp}
\end{align}
where $r_{-2}$ is the characteristic radius (also denoted as $r_{\mathrm{s}}$) of a halo at which the logarithmic slope of the density profile equal to $-2$,
\begin{align}
    \left.\frac{\dd{\log{\qty[\rho_{\mathrm{NFW}} (r)]}}}{\dd{\log{r}}}\right|_{r=r_{-2}} = -2,
\end{align}
and $\rho_{-2} \equiv \rho_{\mathrm{NFW}} (r = r_{-2})$.
The halo concentration $c$ is defined as the ratio between the halo radius (which is adopted as $R_{200\mathrm{c}}$ in this work) and $r_{-2}$,
\begin{align}
    c \equiv \frac{R_{200\mathrm{c}}}{r_{-2}}.
\end{align}
The other NFW parameter $\rho_{-2}$ is related to the concentration as
\begin{align}
    \rho_{-2} = \frac{\rho_{\mathrm{crit}}(a)}{4} \frac{200}{3 \Omega_{\rmm}(a)} \frac{c^3}{f(c)}, 
\end{align}
where $\Omega_{\rmm}(a) \equiv \bar{\rho}_{\rmm}(a) / \rho_{\mathrm{crit}}(a)$ is the matter density parameter at a given scale factor $a$; 
$f(c) \equiv \ln(1+ c) - c / (1 + c)$.

Previously, attention was focused on measuring halo concentrations for the best-fitting \textit{WMAP} or \textit{Planck} cosmologies, or similar models close by in parameter space  \citep[e.g.][]{Gao:2008MNRAS.387..536G,Prada:2012MNRAS.423.3018P,Diemer:2015ApJ...799..108D,Klypin:2016MNRAS.457.4340K,Child:2018ApJ...859...55C}. 
Such calibrated fitting functions cannot be simply extended beyond the cosmological and gravity models for which they have been tested. 
In order to overcome potential problems associated with the  extrapolation of fitting functions to a wider range of cosmologies, we build emulators for the halo concentration-mass relation and halo density profiles.

We study the concentration-mass relation for haloes in the cosmologies covered by the \textsc{FORGE} and \textsc{BRIDGE}\, simulation suites.
We consider only the haloes containing more than $1\,000$ particles, corresponding to a mass of $10^{13}\,h^{-1}M_{\odot}$ for the fiducial \textit{Planck} cosmology.
For several reasons set out below we do {\it not} exclude unrelaxed haloes that contain a large amount of substructure as was done in some previous work \citep[e.g.][]{Neto:2007MNRAS.381.1450N,Prada:2012MNRAS.423.3018P,Klypin:2016MNRAS.457.4340K}.
First, our aim is to predict the halo profile as an ingredient of the halo model, instead of studying the formation and evolution of relaxed haloes.
Galaxies are expected to reside in all haloes, regardless of their dynamic state.
Second, excluding unrelaxed haloes would bias the concentration high because haloes in the rapid mass accretion stage tend to have low concentrations \citep[][]{Child:2018ApJ...859...55C}.
Third, such a cut removes typically $30 \text{-} 50\%$ haloes, which would make the measurement of halo properties less reliable, by introducing larger statistical errors.

To measure halo concentrations, we follow the approach taken by \citet{Mitchell:2019MNRAS.487.1410M} and briefly review the main aspects below.
As mentioned in Section~\ref{subsec:sims}, we use $M_{\mathrm{200c}}$ and $R_{\mathrm{200c}}$ as the definitions of the halo mass and radius, respectively.
The halo centre is adopted as the gravitational potential minimum.
The halo particles are split into $20$ logarithmically spaced radial bins from the halo centre, covering the range $[0.05, 1.00] R_{\mathrm{200c}}$.
We then fit the NFW profile (Eqn.~\eqref{eqn:NFW_profile_exp}) to the density in the radial bins of each halo, treating  the characteristic density and concentration as free variables, by minimising an unweighted $\chi^{2}$,
\begin{align}
    \chi^2 = \sum_i \qty[\log{\rho_{\mathrm{sim}}(r_i)} - \log{\rho_{\mathrm{NFW}} (r_i | c, \rho_{-2})}]^2.
\end{align}
We have checked that weighting the $\chi^2$ function by the number of particles in each radial bin has a negligible impact on the best-fitting concentration values recovered.

The mean value of the best-fitting concentration depends on technical details such as the halo finder used, and the number and range of the radial bins, which would have a non-negligible impact if the NFW profile is not a good fit to the halo profile \citep{Meneghetti:2013arXiv1303.6158M,Dooley:2014ApJ...786...50D}. 
It has been argued that using the median instead of the mean concentration would avoid such non-convergence \citep{Diemer:2015ApJ...799..108D}.
\citet{Neto:2007MNRAS.381.1450N} found that the median of the concentration depends only weakly on the radial range used in the fit, provided that the unrelaxed haloes are removed and $r_{\text{min}} \ge 0.05 R_{\text{vir}}$ in the fitting.

\begin{figure}
    \centering
    \includegraphics[width=\columnwidth]{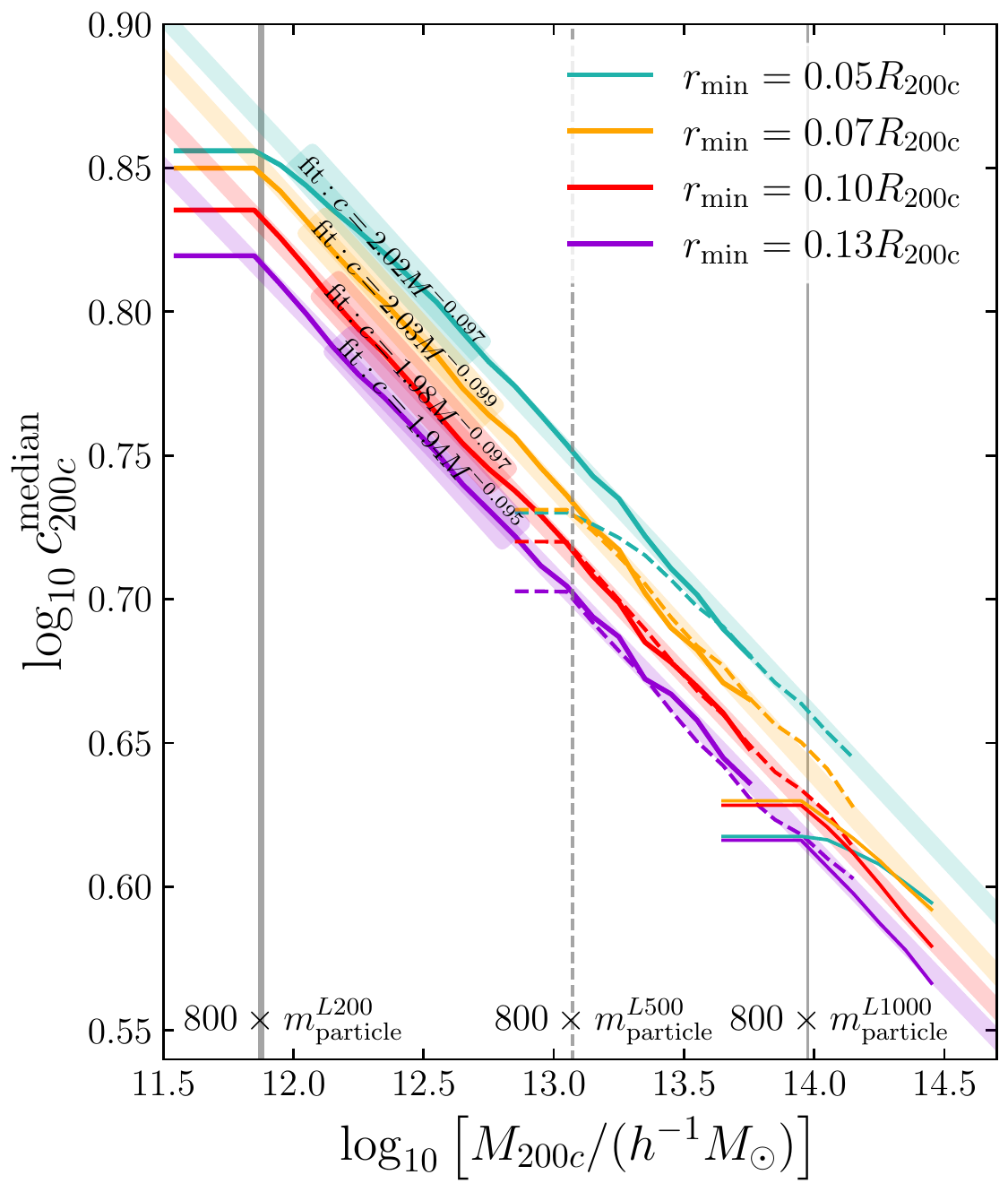}
    \caption[Convergence test of halo concentration measurements.]{Convergence test of halo concentration measurements.
    We fit the NFW profile and measure the halo concentration from three $N$-body simulations with the same particle number and different box sizes, $L200$ (the highest resolution run, thick solid lines), $L500$ (the medium run, thin dashed lines) and $L1000$ (the low resolution run, thin solid lines).
    Colours represent results obtained for different minimum radial ranges in the fitting, as shown by the key, with the maximum radial range fixed at $R_{\mathrm{200c}}$ in each case.
    The vertical lines present the critical mass scales lower than which the concentration measurements are unreliable due to insufficient resolution, which are $\sim800$ times the mass resolutions $m_{\text{particle}}$, as long as $r_{\text{min}} > 0.05 R_{\text{200c}}$.
    The colour-shaded bands show the power fitting ($c(M) = c_0 M^\alpha$) to the measured $c(M)$ relations.
    (\href{https://github.com/chzruan/HM_emulator_paper1_figures/blob/main/conc_Rmin.py}{source code})
    }
    \label{fig:ResolutionTest_all_individual_NFW}
\end{figure}

To test the robustness of the halo concentration measurement, we use three $N$-body simulations from \citet{Mitchell:2021MNRAS.508.4140M} with the same cosmology and particle number, $N_{\text{p}} = 1024^3$, and different box sizes, $L = 200, 500$ and $1000 \, h^{-1} \mathrm{Mpc}$.
We then bin the halo particles, fit the NFW profile and calculate the mean and median values of the concentrations in each mass bin, keeping the maximum radius fixed at $R_{\mathrm{200c}}$ and checking the results for four different $r_{\text{min}}$ values: $(0.05, 0.07, 0.10$ and $0.13) R_{\mathrm{200c}}$.
The results are shown in Fig.~\ref{fig:ResolutionTest_all_individual_NFW}.
All concentration-mass relations are well fitted by a power law,
\begin{align}
    c(M) = c_0 M^{\alpha},
\end{align}
with $c_0$ the amplitude and $\alpha$ the index, as represented by the coloured bands in the figure.
The median concentrations (as well as the mean) and the best-fitting amplitude are still sensitive to the minimum radius, with a relative difference of up to $10$ per cent.
However, the power indices are practically the same for different fitting ranges.

The convergence test also confirms our choice of the minimum particle-number cut applied to define the halo sample used.
The concentrations from the lower resolution simulations start deviating from those of the higher resolution runs around the halo mass corresponding to $\sim 800$ times the particle mass.
This pivot mass also depends weakly on the radial range used to fit the density profile, with smaller minimum scales having larger pivot masses.

\subsection{The averaged halo profile estimated from the halo-mass correlation function}
\label{subsec:halo_profile_ave_xihm}

The normalised halo density profile, $u(r|M)$, that appears in the halo model can be estimated by measuring the halo-mass cross-correlation functions from an $N$-body simulation.
As mentioned in Section~\ref{sec:intro}, the halo model assumes that the matter density field consists of a superposition of haloes at locations $\bdx_i$ with masses $M_i$, so that the matter field can be written as
\begin{align}
    \rho_\rmm (\boldsymbol{x}) &= \sum_i M_i \, u\qty(|\boldsymbol{x} - \boldsymbol{x}_i| \Big| M_i) \\
    &= {\small \int \dd[3]{\boldsymbol{x}'} \int \dd{M}  \bigg[ \sum_i \delta^{\mathrm{(D)}} (M - M_i) \, \delta^{\mathrm{(D)}} (\boldsymbol{x}' - \boldsymbol{x}_i)} \notag \\
    & \phantom{= \small \int \dd[3]{\boldsymbol{x}'} \int \dd{M}}\quad M \, u\qty(|\boldsymbol{x} - \boldsymbol{x}'| \Big| M)\bigg], \label{eqn:rhom_hm_sum_wef}
\end{align}
where
\begin{itemize}
    \item The summation is for all haloes; the same results can be derived by taking the summation over all microcells which are made to be so small that each cell contains no more than one halo centre \citep[e.g.][]{Peebles:1980lssu.book.....P};
    \item $\delta^{\mathrm{(D)}}(\cdots)$ is the Dirac delta function;
    \item $\displaystyle \sum_i \delta^{\mathrm{(D)}} (M - M_i) \, \delta^{\mathrm{(D)}} (\boldsymbol{x}' - \boldsymbol{x}_i) \equiv \frac{\dd{n(\bdx'; M)}}{\dd{M}} $ is the \textit{local} HMF, whose integral over the halo mass gives the halo number density field at the field point $\bdx'$:
    \begin{align}
        \int \dd{M} \frac{\dd{n(\bdx'; M)}}{\dd{M}} = n_{\rmh} (\bdx'),
    \end{align}
    and its ensemble average gives the common dHMF,
    \begin{align}
        \expval{\frac{\dd{n(\bdx'; M)}}{\dd{M}}} &=  \expval{\sum_i \delta^{\mathrm{(D)}} (M - M_i) \, \delta^{\mathrm{(D)}} (\boldsymbol{x}' - \boldsymbol{x}_i)} \notag \\
        &= \dv{n(M)}{M};
    \end{align}
    \item $\displaystyle u\qty(|\boldsymbol{x} - \boldsymbol{x}_i| \Big| M_i) \equiv \frac{\rho_{\mathrm{h}}\qty(|\boldsymbol{x} - \boldsymbol{x}_i| \Big| M_i)}{M_i}$ denotes the density profile of a halo centred at $\boldsymbol{x}_i$, which is also assumed to be spherically symmetric and depends only on its mass,
    \begin{align}
        u \qty(|\boldsymbol{x} - \boldsymbol{x}_i| \big| M_i) = \begin{cases}
            \rho_{\mathrm{h}} \qty(|\boldsymbol{x} - \boldsymbol{x}_i | \Big| M_i) / M_i, & |\boldsymbol{x} - \boldsymbol{x}_i| \le R_{\text{halo}}; \\
            0, & |\boldsymbol{x} - \boldsymbol{x}_i| > R_{\text{halo}};
        \end{cases}
    \end{align}
    and $u$ is normalised: 
    \begin{align}
    \int \dd[3]{\boldsymbol{x}} u\qty(|\boldsymbol{x} - \boldsymbol{x}_i| \big| M_i) &= 1.
    \end{align}
\end{itemize}

For two different populations of objects with overdensity fields $\delta_a (\bdx)$ and $\delta_b (\bdx)$, the two-point cross-correlation function is defined as 
\begin{align}
    \xi_{ab}(\bdr) \equiv \expval{\delta_a(\bdx) \, \delta_b(\bdx + \bdr)}, \label{eqn:xiab_2pcf_def}
\end{align}
where $\expval{\cdots}$ denotes ensemble averaging. 
Assuming statistical isotropy reduces $\xi_{ab}(\bdr)$ to a function of separation only.
In the case of the cross-correlation between halo centres and the matter field, the cross-correlation function is given by
\begin{align}
    \xi_{\rmhm}(r) &\equiv \expval{\delta_{\rmh}(\bdx) \, \delta_{\rmm}(\bdx + \bdr)} \\
    &= \frac{1}{\bar{n}_{\rmh} \bar{\rho}_{\rmm}} \expval{n_{\rmh} (\bdx) \, \rho_{\rmm} (\bdx')} - 1, \label{eqn:xihm_nh_rhom_defdsfvs}
\end{align}
where $\bdx' \equiv \bdx + \bdr$, and the halo number density fluctuation field is defined as
\begin{align}
    \delta_{\rmh}(\bdx) \equiv \frac{n_{\rmh}(\bdx) - \bar{n}_{\rmh}}{\bar{n}_{\rmh}}.
\end{align}

Now we derive the expressions for the halo and matter fields in Eqn.~\eqref{eqn:xihm_nh_rhom_defdsfvs} within the halo model framework.
In realistic $N$-body simulations, halo catalogues always have a finite halo mass range, limited by the simulation force/mass resolution and the box size.
We can define the halo selection function for a given mass range (MR) as
\begin{align}
    \phi(m | \mathrm{MR}) \equiv \begin{cases}
        1, & \text{if}\ m \in \text{MR},\\
        0, & \text{otherwise}. \label{eqn:phi_mr_def_wefds}
    \end{cases}
\end{align}
In practice, MR can be a narrow mass interval, $[M, M+\dd{M}) \approx [M, M + \Delta M)$, or a mass threshold interval, $[M, \infty)$.
For a given halo sample, the corresponding halo number density field can be expressed as 
\begin{align}
    n_{\rmh} (\boldsymbol{x}) = \sum_{i} \delta^{\mathrm{(D)}} (\boldsymbol{x} - \boldsymbol{x}_i) \, \phi(m_i | \mathrm{MR}). \label{eqn:nh_hm_mr_efdvs}
\end{align}
To obtain the expression for $\langle n_{\rmh}(\bdx) \, \rho_{\rmm}(\bdx') \rangle$ in the halo model, we can plug the expressions for the halo and mass fields, Eqns.~\eqref{eqn:rhom_hm_sum_wef} and \eqref{eqn:nh_hm_mr_efdvs}, into Eqn.~\eqref{eqn:xihm_nh_rhom_defdsfvs}.
The full expression can be found in Eqns.~(7) and (8) of \citet{Garcia:2021MNRAS.505.1195G}.
We focus only on the internal structure of the halo and, therefore, on the 1-halo term, which reads
\begin{align}
    \expval{n_{\rmh}(\bdx) \, \rho_{\rmm}(\bdx')}_{\mathrm{1h}} = \int \dd{m} \frac{\dd{n}}{\dd{m}} m \,\phi(m | \mathrm{MR}) \, u(r | m) \\
    = \begin{cases}
        \dd{n(M)} M \, u(r|M), & \mathrm{MR} = [M, M + \dd{M}), \\
        \displaystyle\int_{M}^{\infty} \dd{m} \dndm \, m \, u(r|m), & \mathrm{MR} = [M, \infty),
    \end{cases}
\end{align}
where $\dd{n(M)}$ is the number density of halos in the mass range $[M, M + \dd{M})$.
The 1-halo term of the halo-mass correlation function is
\begin{align}
    \xi_{\rmhm}^{\mathrm{1h}}\qty(r | M) &= \frac{M\, u(r | M)}{\bar{\rho}_{\rmm}} - 1 = \frac{\rho_{\mathrm{h}}(r | M)}{\bar{\rho}_{\rmm}} - 1, \\
    \xi_{\rmhm}^{\mathrm{1h}}\qty(r |\!>\!M) &= \frac{1}{\bar{\rho}_{\rmm} \, \bar{n}_{\rmh}(\!>\!M)} \int_M^{\infty} \dd{m} \dndm \, m \, u(r|m) - 1.
\end{align}

We are interested in the average density profile of haloes in a narrow mass interval $[M,  M + \dd{M})$.
However, the noise level of the simulation measurements for such halo properties and statistics is high because of the low number density.
We bypass this problem by measuring $\xi_{\rmhm}^{\mathrm{1h}}\qty(r | >M)$ (which involves more haloes and therefore is a smoother function) and taking the partial derivative with respect to mass,
\begin{align}
    u(r | M) &=  -\frac{\bar{\rho}_{\mathrm{m}}}{ M} \qty[\frac{\dd{n(M)}}{\dd{M}}]^{-1} \notag \\
    & \quad \times \frac{\partial}{\partial M} \qty{\bar{n}_{\rmh}(>M) \, \Big[1 + \xi_{\text{hm}}^{\mathrm{1h}}\qty(r | >M) \Big]}. \label{eqn:urm_xihm_nm_est}
\end{align}

\section{Using Neural Networks for Regression Problems}
\label{sec:nn}

Given a data set comprised of independent variables (also called \textit{features}) and dependent variables (\textit{labels}), there exists an unknown underlying function mapping the inputs to the outputs.
We can use supervised learning algorithms to approximate this function, which falls in the category of a regression problem.
In the context of structure formation, the features consist of cosmologies $\boldsymbol{\mathcal{C}}$, redshifts $z$, halo masses $M$ or number densities $n_{\rmh}$, etc.
The labels of interest include basic properties of haloes such as halo mass functions, concentration-mass relations and correlation functions.
In the case of structure formation, the ``functions'' between the features and labels are known but expensive: we can run $N$-body simulations given a set of cosmological parameters $\boldsymbol{\mathcal{C}}$, save the snapshot at a given redshift $z$, identify haloes and measure the properties of haloes.
But it is computationally intractable to perform $\gtrsim \mathcal{O}(10^4)$ cosmological simulations to explore the parameter space in a typical MCMC analysis.

As shown in previous works on cosmic emulation, such as \citet{Nishimichi:2019ApJ...884...29N,DeRose:2019ApJ...875...69D.AemulusI,Cuesta-Lazaro:2022arXiv220805218C}, and as we will report in the following sections, it is possible to construct emulators for halo properties by running affordable numbers (e.g., $50 \text{-} 100$) of simulations and interpolating in a  high-dimensional parameter space.
Emulators can give accurate predictions for halo properties for new models without running additional simulations. 
Thanks to the development of algorithms and computing power, statistics and machine learning provide us with a wealth of tools to solve such regression problems.

In a regression analysis, the typical progress of approximating a function can be summarised as:
\begin{itemize}
    \item Define a functional form $\boldsymbol{y} = f(\boldsymbol{x} | \boldsymbol{\theta})$ with adjustable or trainable parameters $\boldsymbol{\theta}$. 
    For example, in the simplest and most common form --- linear regression --- the labels are assumed to be linear combinations of the features and the coefficients are the trainable parameters.
    \item Define a loss function on the training data set $D$ to quantify the difference between the real and predicted values of the target, e.g., the sum of the absolute differences (which reduces the weight of outliers),
    \begin{align*}
        L(\boldsymbol{\theta} | D) \equiv \sum_{(\boldsymbol{x}_i, \boldsymbol{y}_i) \in D} \big|\boldsymbol{y}_i - f(\boldsymbol{x}_i|\boldsymbol{\theta})\big|.
    \end{align*}
    \item Find the optimal parameters ${\boldsymbol{\theta}}_{\star}$ to minimise the loss function (train the model).
\end{itemize}

Gaussian process (GP) regression  \citep[e.g.][]{williams2006gaussian} has been widely adopted in  cosmological emulation projects, such as \textsc{Dark Quest} \citep{Nishimichi:2019ApJ...884...29N}, \textsc{Aemulus} \citep{DeRose:2019ApJ...875...69D.AemulusI}, the Coyote Universe \citep{Heitmann:2010ApJ...715..104H}, the Mira-Titan Universe \citep{Bocquet:2020ApJ...901....5B.MTHMF}, \textsc{Cosmic Emulators} \citep{Kwan:2013ApJ...768..123K} and \textsc{FORGE} \citep{Arnold:2022MNRAS.515.4161A.FORGE_1}.
GP regression is non-parametric, i.e., no specific functional form is assumed.
However, GPs are not easy to apply to large data sets because of their $\mathcal{O}(N^3)$ scaling where $N$ is the size of the training data. 
Therefore, the aforementioned projects usually emulate matter, halo and/or galaxy properties using a combination of principal component analysis, to reduce the dimensionality of the data vector, and GP, to fit the dependence of the principal component coefficients on cosmology.

The machine learning algorithm we adopt is a fully connected neural network \citep[e.g.][]{bishop1995neural,Alom:electronics8030292,2021arXiv210611342Z}. This is a type of artificial neural network in which all neurons in one layer are connected to the neurons in the next layer.
The neural network algorithm has been widely applied to cosmology \citep[e.g.,][]{Agarwal:2012MNRAS.424.1409A,Agarwal:2014MNRAS.439.2102A,Jennings:2019MNRAS.483.2907J,Kobayashi:2020PhRvD.102f3504K,Cuesta-Lazaro:2022arXiv220805218C}, and its performance has been shown to be competitive or sometimes better than other methods.
Neural networks have a strong fitting ability, as reflected by the \textit{universal approximation theorem} \citep{cybenko1989approximation,hornik1989multilayer,goodfellow2016deep}.
However, this theorem does not provide a means to  construct optimised neural networks, but merely guarantees their existence.
Also, this strong fitting ability also makes neural networks more susceptible to the overfitting problem compared to GPs. 
Therefore, we need to carefully check the emulator performance using independent test data sets and tune the network architecture so that the generalisation error is successfully suppressed.
Moreover, we show dimensionality reduction is not necessary when using neural networks, which also improves the accuracy of the emulator predictions.

A neural network is an interconnection of neurons arranged in a series of layers, with each neuron in a layer connected to all other neurons in adjacent layers with different weightings.
One can impart values on to the neurons of the first layer (called the input layer), have a number of hidden layers and finally obtain the output from the last layer.
For example, in this work, the neural networks emulating the halo mass function at fixed redshift have five nodes in the input layer, corresponding to the halo mass and four cosmological parameters, and one node in the last layer outputting the HMF,
\begin{align}
    n\big(>M | \Omega_{\mathrm{m0}}, h, S_8, \log_{10}|\bar{f}_{R0}| \ \text{or}\, \log_{10}|H_0 r_{\mathrmc}|\big).
\end{align}

Neural networks use activation functions to impart non-linearities into the fitting.
\ac{ReLU} \citep{agarap2018deep} is the most commonly used activation function in current neural networks used to add non-linearities in the mapping between inputs and outputs, which is defined as
\begin{equation}
    \mathrm{ReLU}(x) = \max(0, x),
\end{equation}
where $x$ is the output of the previous layer of the neural network. 
Note that the activations of \ac{ReLU} are not differentiable at $x = 0$. 
Here, however, we are interested in functions that are differentiable with respect to their inputs and, in particular, with respect to the cosmological parameters.
Therefore, throughout, we use \ac{GELU} \citep{Hendrycks:2016arXiv160608415H.GELUs} as the activation function instead,
\begin{equation}
    \mathrm{GELU}(x) = 0.5x\qty[1 + \erf\left(\frac{x}{\sqrt{2}}\right)].
\end{equation}

To find the optimal parameters $\boldsymbol{\theta}_{\star}$ that reproduce the halo properties measured in the $N$-body simulations, we minimise the L1-norm loss function,
\begin{equation}
\mathcal{L} = \frac{1}{N} \sum_{i=0}^N |y^i_\mathrm{true} - y^i_\mathrm{predicted}|
\end{equation}
using the Adam optimiser \citep{Kingma:2014arXiv1412.6980K.Adam}.  
Compared to the mean squared error, the loss of L1 reduces the importance given to outlier errors.
To avoid fine-tuning the learning rate, we adopt a learning rate scheduler that reduces the learning rate by a factor of $10$ every time the validation loss does not improve after $30$ epochs. 
We also stop the training process when the validation loss does not improve after $100$ epochs. 
This iterative reduction of the learning rate allows the model to quickly learn the broad characteristics of the data and later reduce the errors with a smaller learning rate. 
The initial learning rate is always set to $0.01$.

\subsection{Ideal emulation tests}
\label{subsec:ideal_emu_test}

To gain a preliminary impression of the emulation process, and to guide the design of emulators in the future, we perform emulation tests under ideal conditions.
The halo properties are generated for a limited number of randomly selected cosmologies using analytical methods or fitting formulae, which are noise-free mappings from cosmologies to halo properties.
Then we use these data to train neural networks and emulate the ``true'' model.
To evaluate the performance of the emulator, we compare the true values with the emulator predictions using independent test data sets.

\begin{figure}
    \centering
    \includegraphics[width=\columnwidth]{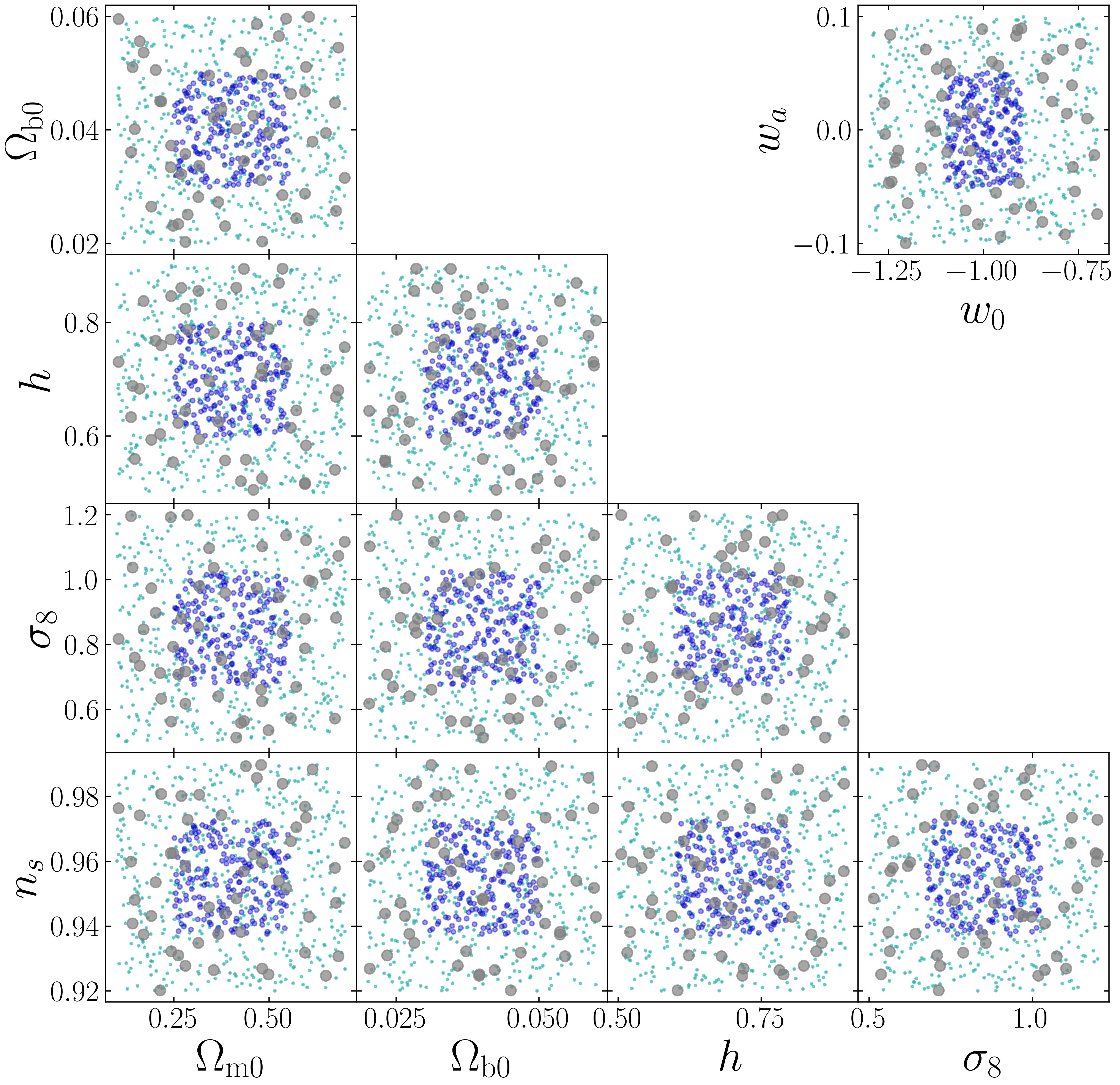}
    \caption{Visualisation of the seven-parameter $(w_0\text{-}w_a)$CDM cosmologies studied in the ideal emulation tests.
    {Grey} dots show the {training} set including $50$ nodes. 
    {Green} dots represent full-range {test} set consisting of $500$ nodes covering the same range as that of the training set. 
    {Blue} dots show the half-range test set including $500$ nodes in the inner half region of the parameter space. (\href{https://github.com/chzruan/HM_emulator_paper1_figures/blob/main/nodes_ideal_tests.py}{source code})}
    \label{fig:NodeParams_ideal_emu}
\end{figure}

The cosmologies of the training set cover $50$ flat geometry $(w_0 \text{-} w_a) $ CDM models \citep{Linder:2003PhRvL..90i1301L}, where the equation-of-state parameter for dark energy is parameterised in terms of the expansion factor, $a$, as 
\begin{align}
    w(a) = w_0 + w_a (1 - a).
\end{align}
A key aspect of building emulators is an efficient sampling scheme.
As the training dataset, the $50$ cosmologies were sampled using optimal minimax distance sliced Latin hypercube designs \citep{Ba:doi:10.1080/00401706.2014.957867}  in a   seven-dimensional cosmological parameter space, 
\begin{align}
    \boldsymbol{\mathcal{C}} = \Big\{ \Omega_{\rmm0}, \Omega_{\mathrm{b0}}, h, \sigma_8, n_{\mathrms}, w_0, w_a \Big\},
\end{align}
as shown by the grey dots in Fig.~\ref{fig:NodeParams_ideal_emu}.
The range of parameters is
\begin{equation}
\begin{split}
    0.1 &< \Omega_{\rmm0} < 0.7,\quad 0.02 < \Omega_{\mathrm{b0}} < 0.06, \\
    0.5 &< h < 0.9, \quad 0.5 < \sigma_8 < 1.2, \\
    0.92 &< n_{\mathrms} < 0.99, \\
    -1.3 &< w_0 < -0.7,\quad -0.1 < w_a < 0.1,
\end{split}
\end{equation}
while the upper and lower parameter limits depart  significantly from the current best-fitting $\Lambda$CDM background cosmology from the \textit{Planck} satellite \citep{Planck2018}.
We also generate two test data sets that were not used in the training: both consist of $500$ random cosmologies; one set covers the same parameter range as that of the training set, and the other one covers the inner half-region (in terms of the length per dimension, instead of volume) of the parameter space.
The cosmologies in the full- and half-range test data sets are shown in green and blue dots in  Fig.~\ref{fig:NodeParams_ideal_emu}, respectively.

We choose to emulate two basic properties of haloes in the tests: the concentration-mass relation $c(M)$, and the cumulative HMF $\bar{n}_{\rmh}(>M)$.
For given cosmologies, we generate the $c(M)$ relation calibrated in \citet{Prada:2012MNRAS.423.3018P}, using the publicly available Python toolkit \href{https://bdiemer.bitbucket.io/colossus/}{COLOSSUS} \citep{Diemer:2018ApJS..239...35D.COLOSSUS}, and compute the \citet{Tinker:2008ApJ...688..709T.HMF} HMF with the Python package  \href{https://hmf.readthedocs.io/en/latest/_autosummary/fitting_functions/hmf.mass_function.fitting_functions.Tinker08.html}{\texttt{hmf}} \citep{Murray:2013A&C.....3...23M.HMFcalc,Murray2021A&C....3600487M.THEHALOMOD}.
As described in Section~\ref{subsec:hmf_data} and Fig.~\ref{fig:HMF_emu_pipeline}, we train the emulators directly on the ratio of the cumulative HMF between the target HMF and a fitting formula to reduce the dynamic range and improve the interpolation accuracy.
We choose the HMF calibrated in \citet{Jenkins:2001MNRAS.321..372J} as the reference.

\begin{figure*}
    \centering
    \includegraphics[height=0.4\textwidth]{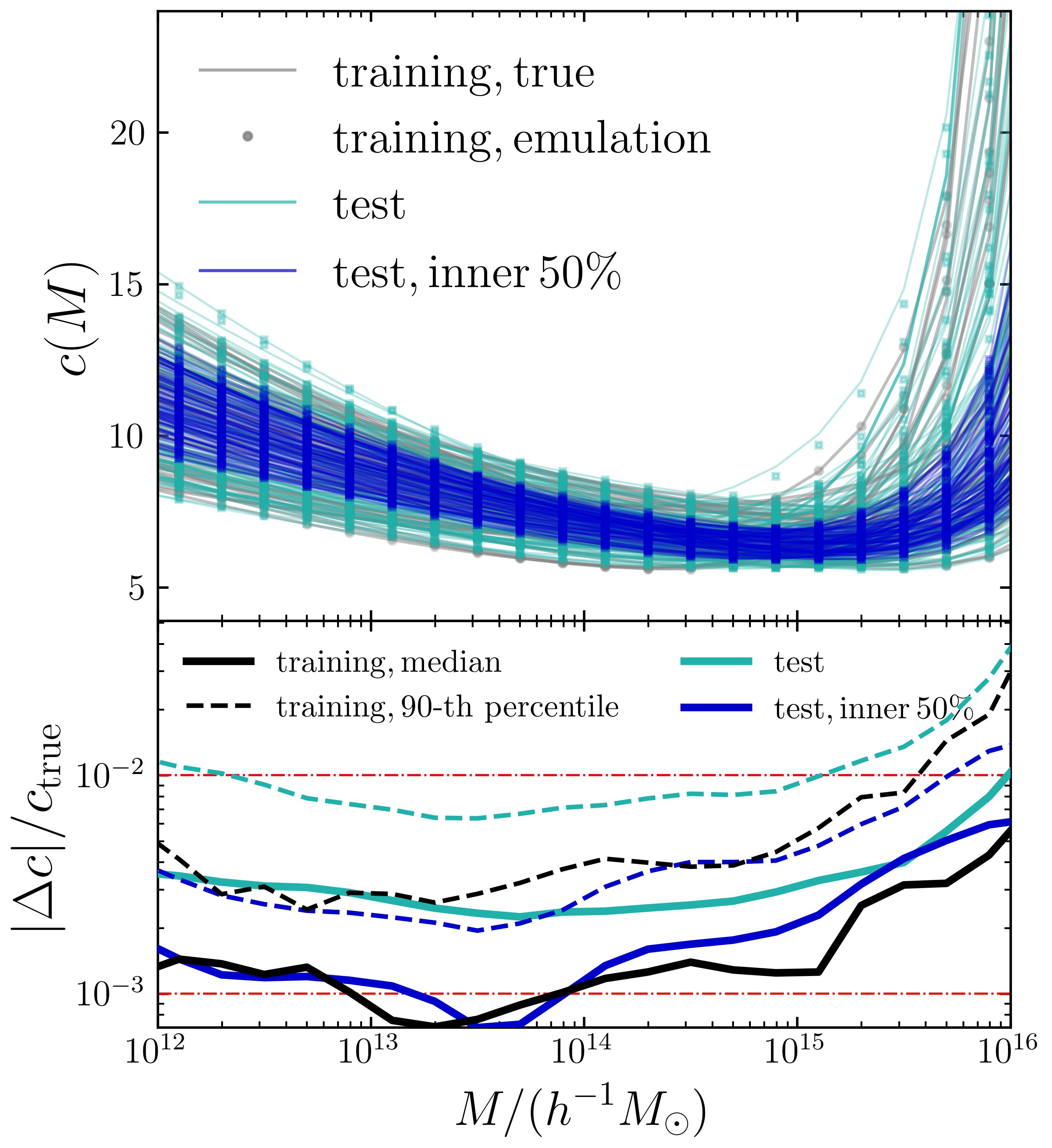}%
    \includegraphics[height=0.4\textwidth]{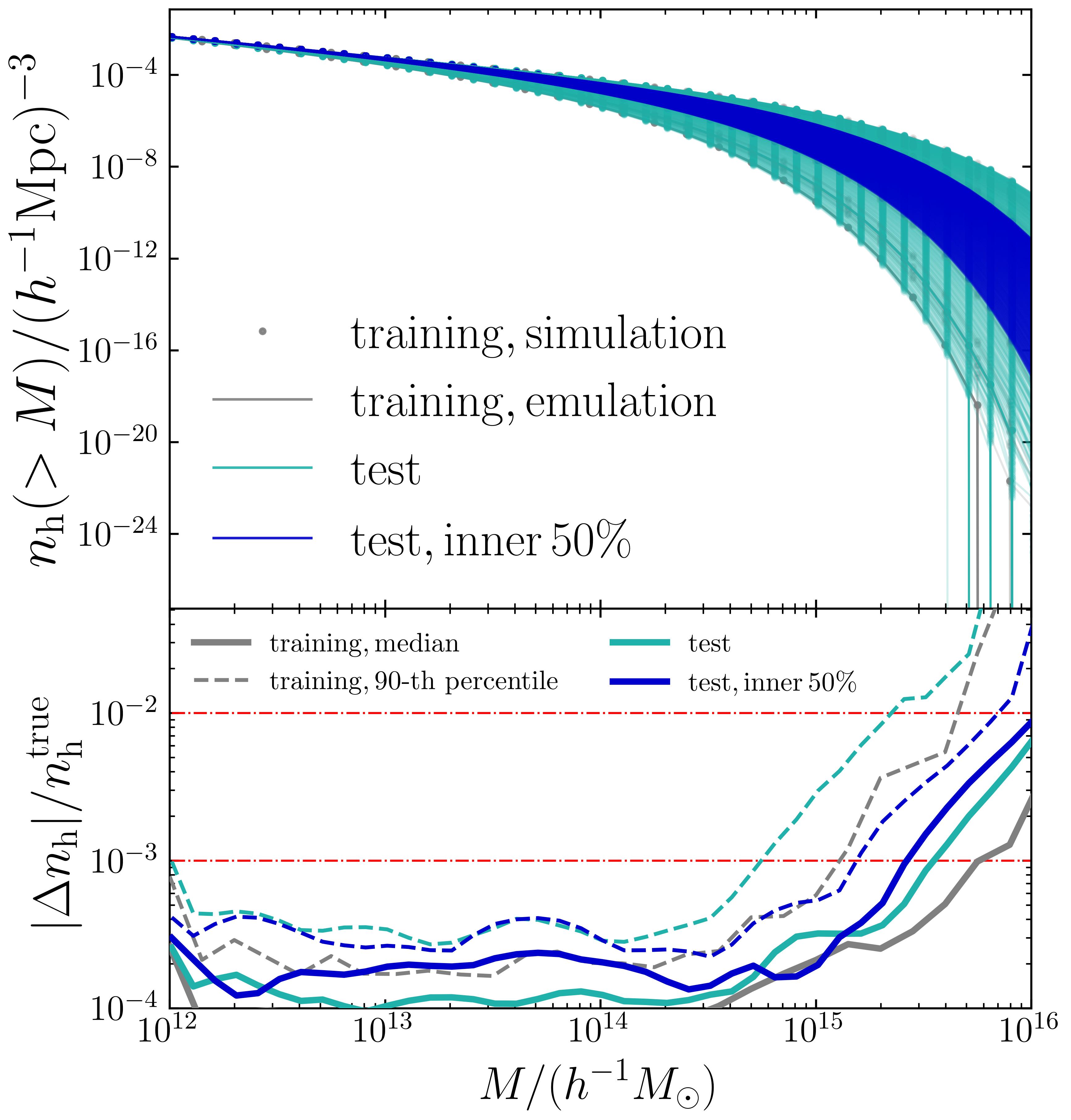}
    \caption{Ideal emulation tests.
    \textit{Top} panels: The halo concentration-mass relation (left) and halo mass function (right) of the training ({grey}), full-range test ({green}) and half-range test ({blue}) sets, from the analytical methods (dots) and emulators (lines).
    \textit{Bottom} panels: The absolute value of the relative difference between the models and emulators.
    The solid and dashed lines present the median and 90-th percentile of the emulation fractional errors among  the training (black), test (green) and half-range test (blue) sets, which include $50$, $500$ and $500$ cosmologies, respectively. (\href{https://github.com/chzruan/HM_emulator_paper1_figures/blob/main/conc_ideal_emulation.py}{source code~1}, \href{https://github.com/chzruan/HM_emulator_paper1_figures/blob/main/hmf_ideal_emulation.py}{2})
    }
    \label{fig:concentrationEmuVsTruePlot_50nodes_5folds}
\end{figure*}

The upper panels of Fig.~\ref{fig:concentrationEmuVsTruePlot_50nodes_5folds} show the halo properties calculated using the fitting formulae and  emulators.
The fractional errors are shown in the lower sub-panels.
The results show that the emulator reproduces the analytical halo $c(M)$ relation with a sub-percent error in the 7D parameter space with only $50$ training models. 
The performance of the HMF emulator is even better than that of the concentration emulator, although the HMF data span $\sim 20$ orders of magnitude.
The median absolute error of the emulator predictions is lower than $1$ per cent for halo masses $M \lesssim 10^{16}\,h^{-1}M_{\odot}$.

We also note that the emulator precision is generally different at the edge and centre of the parameter space, as revealed by the green and blue lines in the bottom panels of Fig.~\ref{fig:concentrationEmuVsTruePlot_50nodes_5folds}.
This suggests that we should design the parameter space to be wider than the existing cosmological constraints.
The parameter space in our ideal tests is designed to be wide enough that covers some extreme cosmologies, such as $\Omega_{\text{m0}} = 0.7$ and $h = 0.5$.

In general, ideal emulation tests show that under noise-free conditions, neural network emulators can provide accurate interpolations in high-dimensional parameter space, using $50$ efficiently sampled models as the training set.
In the next section, we will present the cosmic emulation in the real situation: the data are measured from simulations and, therefore, are influenced by sample variance and noise.

\begin{table*}
\centering
\caption{Summary of the neural network configurations for emulating the halo properties.
The architecture of a neural network is specified from the input to output layer as $(N_{\text{input}}, N_{\text{hidden1}}, N_{\text{hidden2}}, \cdots, N_{\text{output}})$ with $N$ the number of neurons in each layer. 
}
\label{tab:nn_config}
    \begin{tabular}{ccccc}
    \toprule
     \makecell{Halo \\ Property} & Feature & Label & \makecell{Neural Network \\ Architecture} & Activation \\ 
     \midrule
     HMF & $\boldsymbol{\mathcal{C}}, M_{\mathrm{200c}}$ & Equation~\eqref{eqn:hmf_ratio} & $(5, 64, 32, 1)$ & GELU \\ 
     Concentration & $\boldsymbol{\mathcal{C}}, M_{\mathrm{200c}}$ & $c_{200\mathrm{c}}$ & $(5, 32, 16, 1)$  & GELU \\ 
     $\xi_{\text{hm}}$ & $\boldsymbol{\mathcal{C}}, n_{\mathrm{h}}$ & $\{ r_i^2 \xi_{\text{hm}}(r_i) \}_{i=1}^{N=30}$ & $(5, 128, 32, 30)$  & GELU \\ 
     \bottomrule
    \end{tabular}
\end{table*}

\section{Results}
\label{sec:results}

In this section we demonstrate the ability of a fully connected neural network to reproduce the halo properties measured from \textsc{FORGE} and \textsc{BRIDGE} simulations.
We train different emulators for each gravity theory: $\Lambda$CDM, $f(R)$ gravity and DGP.
The configurations of the neural networks for emulating three halo properties are summarised in Table~\ref{tab:nn_config}.

\subsection{The emulator for the halo mass function}
\label{subsec:hmf_emulator}

\begin{figure*}
    \centering
    \includegraphics[width=\textwidth]{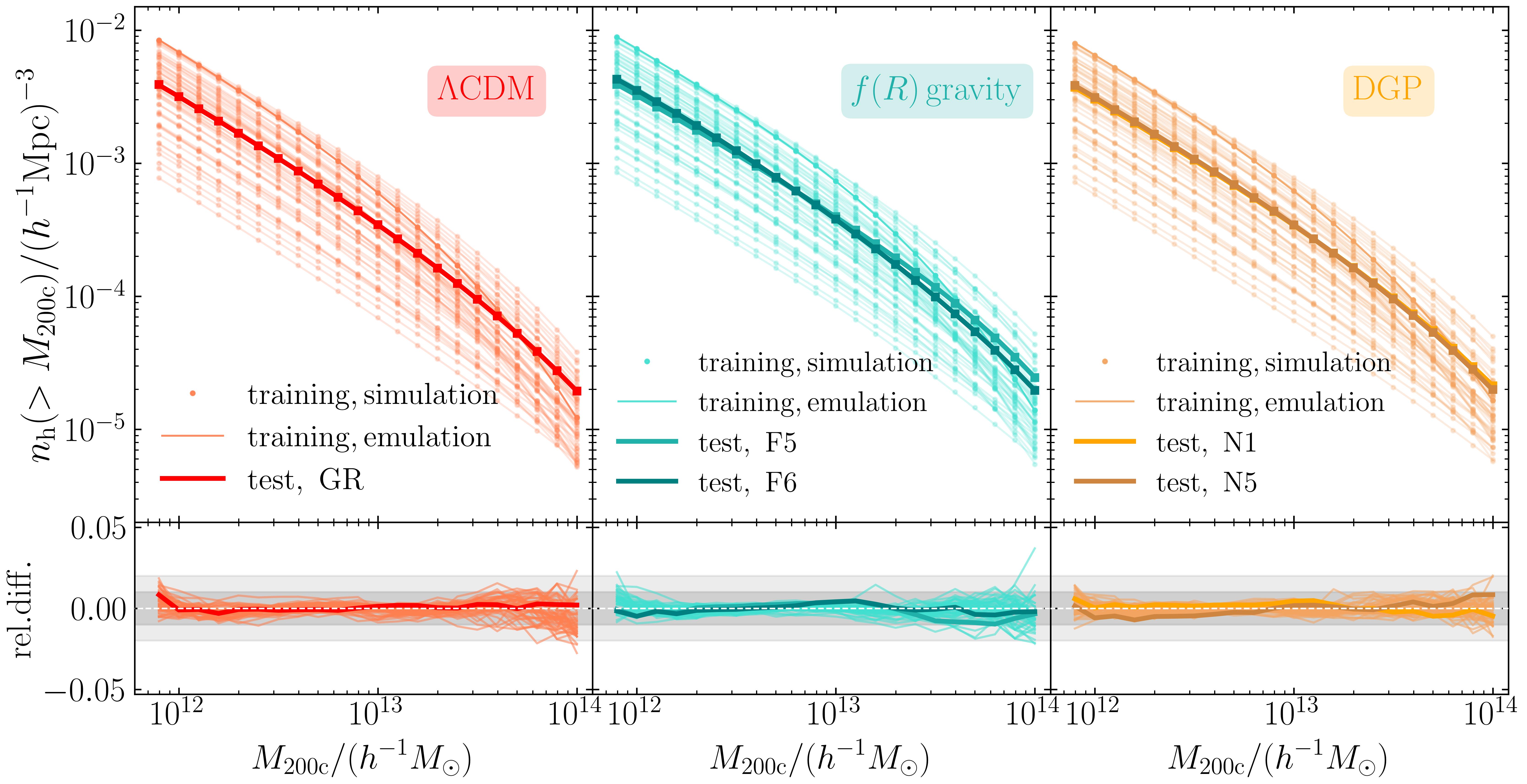}
    \includegraphics[width=\textwidth]{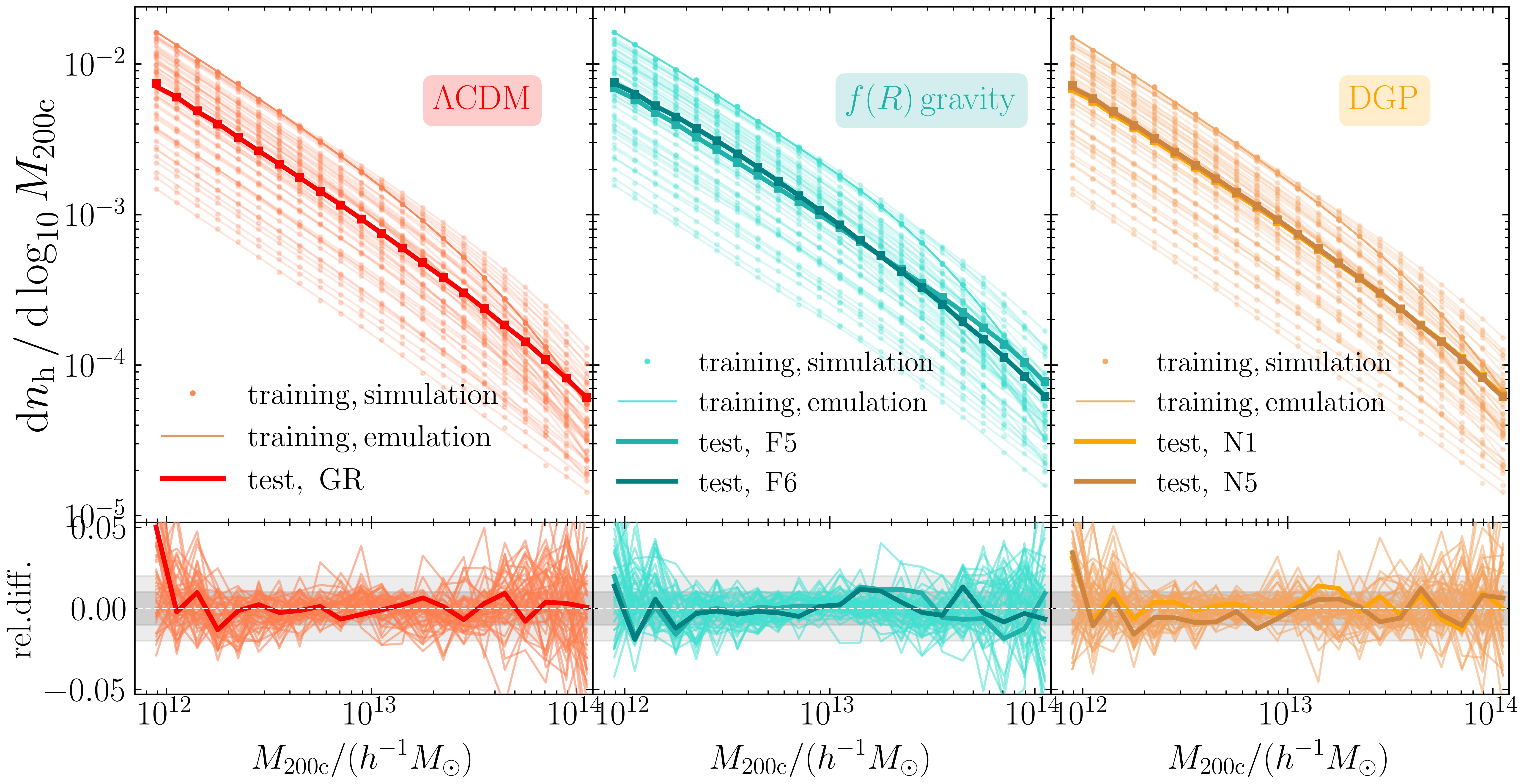}
    \caption{
    Cumulative (the first row) and differential (the second row) halo mass functions from simulation measurements and emulator predictions at $z=0$, for $\Lambda$CDM (left), $f(R)$ gravity (middle) and DGP (right).
    In each panel, the thin lines show the results of the $49$ cosmologies in the training data set, and the thick lines represent those of the test models which were not used in the construction of the emulators.
    In the lower sub-panel, we compare the relative differences between simulations and emulators.
    The dark and light grey bands denote $\pm 1\%$- and $\pm2\%$-level errors, respectively.
    The differential halo mass functions are obtained by finite difference of the cumulative HMFs according to Eqn.~\eqref{eqn:dHMF_fin_diff}.
    (\href{https://github.com/chzruan/HM_emulator_paper1_figures/blob/main/chmf_emu_vs_sim.py}{source code~1}, \href{https://github.com/chzruan/HM_emulator_paper1_figures/blob/main/dhmf_emu_vs_sim.py}{2})
    }
    \label{fig:HMF_emu}
\end{figure*}

As discussed in Section~\ref{subsec:hmf_data}, we train the neural network emulators directly on the ratio of the cumulative HMF between simulation measurements and the fitting formula calibrated in \citet{Tinker:2008ApJ...688..709T.HMF} to reduce the dynamic range and improve the emulation accuracy.
Fig.~\ref{fig:HMF_emu} compares the cumulative and the differential HMFs  from simulations and emulators at $z = 0$, for the $\Lambda$CDM, $f(R)$ gravity and DGP models.
The differential HMF of a given mass bin centred on $\log{M_i}$ is obtained by 
\begin{align}
    \left.\dv{n(M)}{\log{M}}\right|_{\log{M_i}} \approx \frac{ n(> \log{M_i} - \frac{\mathrm{bw}}{2}) - n(> \log{M_i} + \frac{\mathrm{bw}}{2}) }{\mathrm{bw}}, \label{eqn:dHMF_fin_diff} 
\end{align}
where $\mathrm{bw} \equiv \log{M_{i+1}} - \log{M_i}$ is the bin width.
The lower sub-panels show the fractional difference between the emulator prediction and the measured HMFs in a given mass bin,
\begin{align}
    \frac{{\mathrm{HMF}}_{\text{emu}} - {\mathrm{HMF}}_{\text{sim}}}{{\mathrm{HMF}}_{\text{sim}}}.
\end{align}
The performance of the emulator on the training data set is shown by the thin lines of Fig.~\ref{fig:HMF_emu}.
The emulator achieves sub-percent accuracy in reproducing most of the cumulative HMF for halo masses between $10^{12}$ and $10^{14}\,h^{-1}M_{\odot}$.
The residuals of the differential HMF obtained using Eqn.~\eqref{eqn:dHMF_fin_diff} are slightly larger but still show $\lesssim 2$ per cent scatter around zero, with a mean consistent with zero.
The thick lines in Fig.~\ref{fig:HMF_emu} show the emulator predictions for three test models which are not used in the training, as described in Section~\ref{subsec:sims}.
We again find percent-level agreement between the emulator predictions and simulation results.
Furthermore, the fluctuations of the residuals in the test set are much smaller than those of the training models, since each test cosmology has $8$ realisations and the sample variance of the measured dHMFs is suppressed.
This confirms that the errors of the emulator predictions are mainly random instead of systematic.

\subsection{The emulator for the concentration-mass relation}
\label{subsec:cm_emulator}

\begin{figure}
    \centering
    \includegraphics[width=\columnwidth]{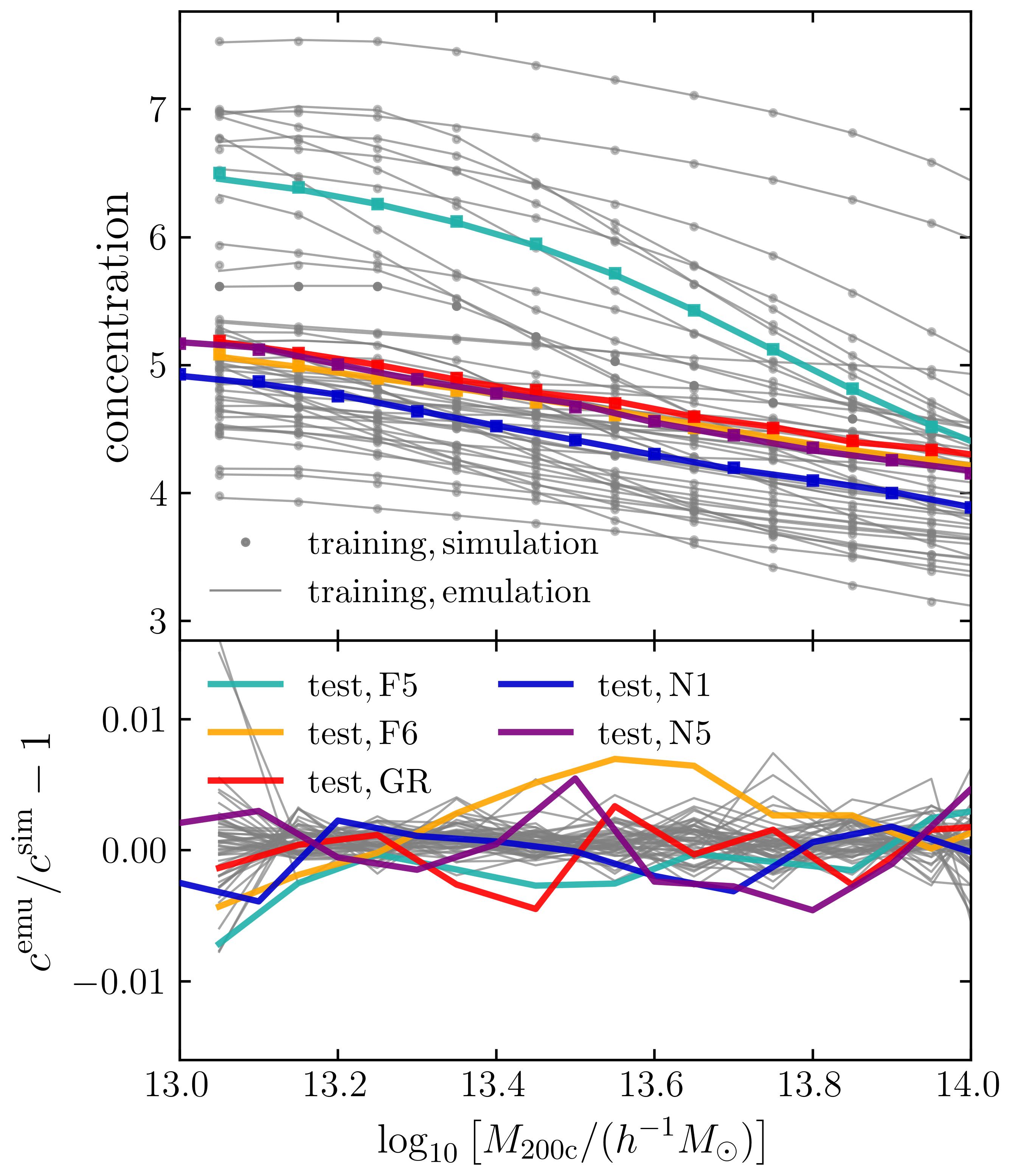}
    \caption{
    Comparison between the halo concentration-mass relations from simulations (points) and emulators (lines), for the $49$ $f(R)$ gravity cosmologies in the training set ({grey}) and test models F5 ({green}), F6 ({orange}), N1 (blue), N5 (purple) and the fiducial \textit{Planck} cosmology (denoted as GR, in {red}).
    The results for the $\Lambda$CDM and DGP training sets are similar to those for $f(R)$ gravity and are therefore not shown here. 
    The lower sub-panel shows the relative difference of the halo concentration between simulations and emulators. 
    The sub-percent differences between the emulator and simulation results are much smaller than the differences between the results for different cosmologies.
    (\href{https://github.com/chzruan/HM_emulator_paper1_figures/blob/main/conc_emu_vs_sim.py}{source code})
    }
    \label{fig:emu_vs_sim_50Nodes_redshift0}
\end{figure}

As shown in Section~\ref{subsec:cm_emulator} and Fig.~\ref{fig:ResolutionTest_all_individual_NFW}, the halo concentrations measured from simulations are sensitive to the radial range used in the fitting, which indicates that this is not the optimal way to describe the halo density profiles.
However, the power law index is not sensitive to the range adopted, which indicates that we can treat the amplitude of the concentration-mass relation as a free parameter to take into account this variation.
We build an emulator for the $c(M)$ relation taking $r_{\text{min}} = 0.10R_{\mathrm{200c}}$ as a representative value. 
The performance of the emulator at $z = 0$ is shown in Fig.~\ref{fig:emu_vs_sim_50Nodes_redshift0}.
The fractional errors are sub-percent for most of the cosmologies in the training and test data sets.

\subsection{The emulator for the halo-mass cross-correlation function}
\label{subsec:rhoh_from_xihm}

As discussed in Section~\ref{subsec:halo_profile_ave_xihm}, the average halo profile can be estimated from the halo-mass cross-correlation function.
The halo profile $u(r|M)$ is directly related to the matter density field cross-correlated with the halo sample in a narrow mass range $[M, M + \Delta M]$.
However, such correlation functions measured from simulations would be rather noisy because of the low halo number density.
To feed the neural networks with smoother data, we measure the cross-correlation functions between the matter field and the halo samples with fixed number densities, $\xi_{\rmhm}(r | \boldsymbol{\mathcal{C}}, \bar{n}_{\rmh})$.
We then use the HMF emulator to translate $\xi_{\rmhm}(r | \boldsymbol{\mathcal{C}}, \bar{n}_{\rmh})$ as a function of number density into $\xi_{\rmhm}(r | \boldsymbol{\mathcal{C}}, M)$ as a function of mass, according to  Equation~\eqref{eqn:urm_xihm_nm_est}.

\begin{figure*}
    \centering
    \includegraphics[height=0.55\textwidth]{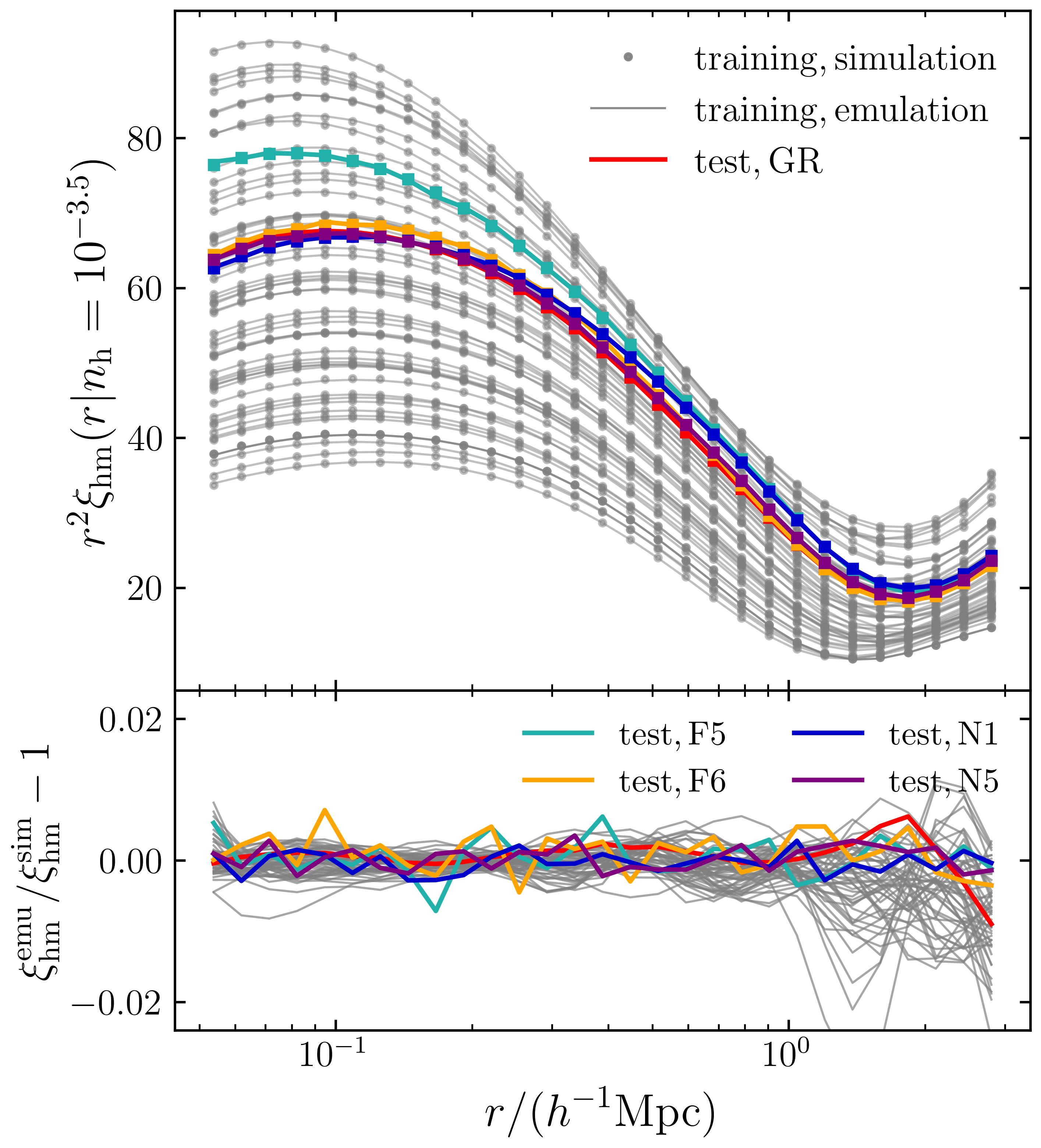}%
    \includegraphics[height=0.55\textwidth]{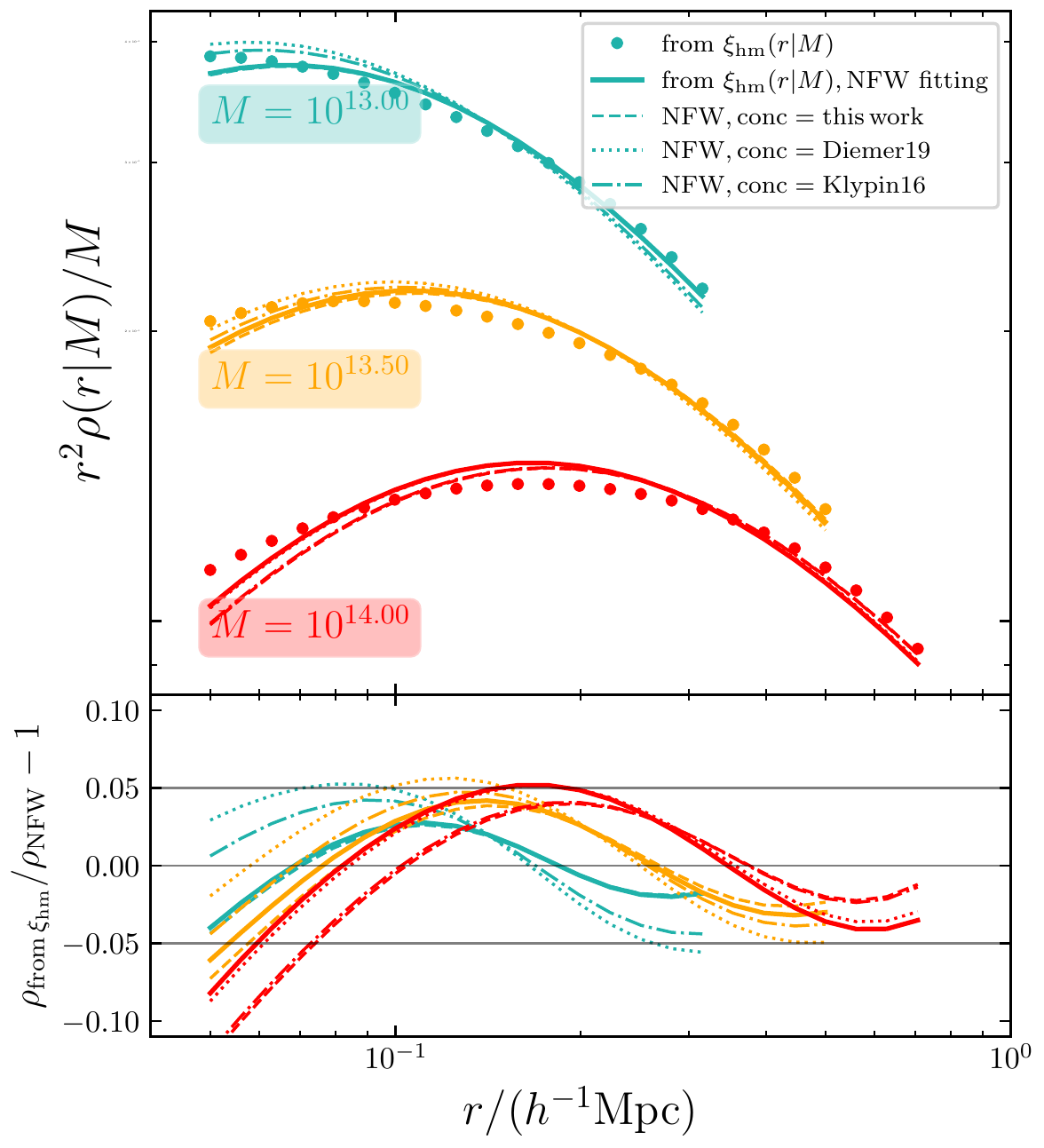}
    \caption{\textit{Left} panel: Halo-mass cross-correlation functions from simulations and emulators at $z=0$, for the $49$ $\Lambda$CDM gravity cosmologies in the training set ({grey}), five test models F5 ({cyan}), F6 ({orange}), N1 (blue), N5 (purple) and the fiducial \textit{Planck} $\Lambda$CDM model ({red}). 
    In the lower sub-panel, we compare the relative difference between the simulations and emulators.
    \textit{Right} panel: Comparison of the normalised halo density profiles, $\rho(r|M) / M$, truncated at $r = R_{\mathrm{200c}}$, for the fiducial \textit{Planck} $\Lambda$CDM model (node 0) at $z=0$.
    The average halo profiles estimated from $\xi_{\text{hm}}(r)$ according to  Equation~\eqref{eqn:urm_xihm_nm_est} are represented by {points}.
    The {solid} lines show the fits to an NFW profile.
    The {dashed}, {dotted} and {dash-dotted} lines show the NFW profiles with three different concentration-mass relations: this work, \citet{Diemer:2019ApJ...871..168D} and \citet{Klypin:2016MNRAS.457.4340K}.
    Colours denote different halo masses.
    The lower sub-panel shows the relative difference between the NFW profiles (lines) and the average profiles (points).
    (\href{https://github.com/chzruan/HM_emulator_paper1_figures/blob/main/xihm_emu_vs_sim.py}{source code})
    }
    \label{fig:profile_NFW_emu_vs_sim_FORGE50}
\end{figure*}

We measure $\xi_{\rmhm}(r | \boldsymbol{\mathcal{C}}, \bar{n}_{\rmh})$ using the high-performance code \href{https://corrfunc.readthedocs.io/en/master/}{\texttt{Corrfunc}} \citep{Sinha:2020MNRAS.491.3022S.corrfunc} for the halo number densities in logarithmically spaced bins over the range
\begin{align}
    \log_{10}\qty[\frac{\bar{n}_{\rmh}}{(h^{-1} \mathrm{Mpc})^{-3}}] = [-5.1, -2.9],
\end{align}
using a bin width of $\Delta \log_{10}\bar{n}_{\rmh} = 0.05$.
The separation $r$ is split into $30$ logarithmically-spaced bins from $0.05\,h^{-1}\mathrm{Mpc}$ (three times the force resolution) to $3\,h^{-1}\mathrm{Mpc}$.
Furthermore, to reduce the dynamic range of the data vector, we opt to emulate $r^2 \xi_{\rmhm}(r)$ instead of $\xi_{\rmhm}(r)$ itself.
The upper-left panel of Fig.~\ref{fig:profile_NFW_emu_vs_sim_FORGE50} shows $\xi_{\rmhm}\big(r | \bar{n}_{\rmh} = 10^{-3.5}\,(h^{-1}\,\mathrm{Mpc})^{-3}\big)$ at $z=0$ for the $49$ $\Lambda$CDM gravity cosmologies along with the test models.

The average halo profile is only related to the 1-halo term of $\xi_{\rmhm}$.
The halo-mass correlation enters the transition between 1- and 2-halo terms as the scale increases.
To estimate the range of the one-halo term, we only consider the scales below $R_{\mathrm{200c}}$, which is related to the adopted mass definition $M_{\mathrm{200c}}$ as
\begin{align}
    M_{\mathrm{200c}}(z) = \frac{4\pi}{3} (R_{\mathrm{200c}})^3 200 \rho_{\mathrm{crit}}(z).
\end{align}

We then build emulators for $\xi_{\rmhm}(r | \boldsymbol{\mathcal{C}}, \bar{n}_{\rmh})$ at each redshift, to reduce the number of features and minimise emulation errors.
The lower left sub-panel of Fig.~\ref{fig:profile_NFW_emu_vs_sim_FORGE50} shows the fractional difference between the simulation measurements and the emulator predictions, in the training set along with the test models.
The emulator achieves sub-percent accuracy for the both the training and the test models.

The average halo density profile can be estimated from $\xi_{\text{hm}}$ using Eqn.~\eqref{eqn:urm_xihm_nm_est}.
In the right panel of Fig.~\ref{fig:profile_NFW_emu_vs_sim_FORGE50}, we compare this type of profile with the NFW profiles combined with three concentration-mass relations from this work, \citet{Klypin:2016MNRAS.457.4340K} and \citet{Diemer:2019ApJ...871..168D}, in five halo mass bins.
We also fit the average profile with an NFW form, using the the data over the range of $[0.1, 1.0]R_{\text{200c}}$.

In the right sub-panel of Fig.~\ref{fig:profile_NFW_emu_vs_sim_FORGE50}, we check the relative difference between the average profiles measured from the simulations and the NFW fits.
There is a $\sim 5\%$ discrepancy between the two types of profiles, regardless of the concentration-mass relation, which shows that the differences between the NFW profiles with different $c(M)$ relations are small.
This level of difference is consistent with the results discovered in Section~5.1.2 of \citet{Nishimichi:2019ApJ...884...29N}.

\section{Emulator Applications}
\label{sec:applications}
With the emulators for the halo mass function and density profile as ingredients, we are able to predict galaxy clustering statistics using the halo model framework \citep{Cooray:2002PhR...372....1C}, and therefore fit galaxy clustering measurements in the joint parameter space of cosmology and a galaxy-halo connection model.
In this section, we demonstrate that the emulator-based halo model reproduces the signals measured from the mock HOD catalogues generated with the same specifications, such as the cosmology, HOD prescription, satellite profile and/or concentration-mass relation.


\subsection{Galaxy two-point correlation function}
\label{subsec:application_xi_gg}
We adopt the halo occupation distribution (HOD) \citep[e.g.][]{Zheng:2005ApJ...633..791Z.HOD} prescription to model the average number of galaxies in a halo as a function of halo mass.
The occupation of central galaxies is parameterised as a Bernoulli distribution, whereas that of satellites is a Poisson distribution \citep{Zheng:2005ApJ...633..791Z.HOD}. 
Both distributions are described by their mean occupation number,
\begin{align}
    \expval{N_{\rmg}}(M) = \langle N_{\mathrmc} \rangle(M) + \langle N_{\mathrms} \rangle(M).
\end{align}
The galaxy number density $\bar{n}_{\rmg}$ can then be obtained by integrating the HMF weighted by the mean occupation,
\begin{align}
    \bar{n}_{\rmg} = \int\dd{M}\dv{n(M)}{M} \big[\expval{N_{\mathrmc}}\!(M)\, + \expval{N_{\mathrms}}\!(M) \big].
\end{align}

Following \citet{Cuesta-Lazaro:2022arXiv220805218C}, we adopt the HOD model in \citet{Zheng:2007ApJ...667..760Z} by introducing the following HOD parameters,
\begin{align}
    \boldsymbol{\mathcal{G}} = \Big\{\underbrace{M_{\text{min}}, \sigma_{\log{M}}}_{\displaystyle\boldsymbol{\mathcal{G}}_{\text{cen}}}, \underbrace{M_1, \kappa, \alpha}_{\displaystyle\boldsymbol{\mathcal{G}}_{\text{sat}}} \Big\}.
\end{align}
The mean occupation number for central galaxies is given by
\begin{align}
    \expval{N_{\mathrmc}}\!(M | {\boldsymbol{\mathcal{G}}}) = \frac{1}{2} \qty[1 + \erf\qty(\frac{ \log{M} - \log{M_{\text{min}}} }{ \sigma_{\log{M}} })].
\end{align}
The mean central HOD, $\expval{N_{\mathrmc}}\!(M)\,$, can be interpreted as the probability that a halo with mass $M$ hosts a central galaxy. 
The mean central HOD considered here has the asymptotic behaviour that  $\expval{N_{\mathrmc}} \to 0$ for haloes with $M \ll M_{\text{min}}$, while $\expval{N_{\mathrmc}} \to 1$ for haloes with $M \gg M_{\text{min}}$.

The mean satellite HOD is parameterised as
\begin{equation}
    \expval{N_{\mathrms}}\!(M | {\boldsymbol{\mathcal{G}}}) = \expval{N_{\mathrmc}}\!(M | {\boldsymbol{\mathcal{G}}})  \lambda_{\mathrms} (M),
\end{equation}
where
\begin{equation}    
    \lambda_{\mathrms} (M) = \qty(\frac{M - \kappa M_{\text{min}}}{M_1})^{\alpha}.
\end{equation}
Following the commonly-used prescription, we assume that satellite galaxies reside only in a halo that already hosts a central galaxy. 
Hence, in the above equation, satellite galaxies can only reside in haloes with $\expval{N_{\mathrmc}} = 1$. 
Then we assume that the number distribution of satellite galaxies in a given host halo follows the Poisson distribution with mean $\lambda_{\mathrms}(M)$:
\begin{align}
    P(N_{\mathrms} | N_{\mathrmc} = 1) &=  \frac{\qty[\lambda_{\mathrms}(M)]^{N_{\mathrms}} \mathrm{e}^{ -\lambda_{\mathrms}(M)}}{N_{\mathrms}!},  \\
    \intertext{and} 
    P(N_{\mathrms} | N_{\mathrmc} = 0) &= \delta^{\mathrm{Kr}}_{N_{\mathrms},0},
\end{align}
where $\delta^{\mathrm{Kr}}_{ij}$ stands for the Kronecker delta.

Given the HOD model, we populate dark matter haloes in $8$ test simulations of the fiducial \textit{Planck} cosmology with mock galaxies and measure the galaxy clustering signals, using the following randomly selected HOD parameters:
\begin{equation}
\begin{split}
    & \log{[M_{\text{min}} / (h^{-1} M_{\odot})]} = 12.5, \ \sigma_{\log{M}} = 0.6915, \\
    &\kappa = 0.51, \log{[M_1 / (h^{-1} M_{\odot})]} = 12.9, \ \text{and}\ \alpha = 0.9168.
\end{split}
\end{equation}

We can also express the galaxy \ac{TPCF} in terms of dark matter halo properties in the halo model framework.
First, we split the one- and two-halo terms into correlations of central and satellite galaxies as
\begin{align}
    \xi_{\rmgg}(r) &= \xi_{\mathrm{cs}}^{\mathrm{1h}}(r) + \xi_{\mathrm{ss}}^{\mathrm{1h}}(r) \notag \\
    &\phantom{=} + \xi_{\mathrm{cc}}^{\mathrm{2h}}(r) + \xi_{\mathrm{cs}}^{\mathrm{2h}}(r) + \xi_{\mathrm{ss}}^{\mathrm{2h}}(r).
\end{align}
The terms involving both centrals and satellites lead to a convolution of the halo profiles and/or the halo \ac{TPCF}, following  $\displaystyle \int\dd[3]{\bdx} u(x|M) \, u\qty(|\bdx + \boldsymbol{r}| \Big| M)$.
It is therefore more convenient to compute these terms in Fourier space, where convolutions in coordinate space become simple products of the Fourier modes.
Here, we focus on the one-halo term only. The two-halo term involving the emulation of halo clustering will be the topic of the subsequent papers in this series.
The expressions for the 1-halo term of the galaxy \ac{TPCF} after the central-satellite split are given by
\begin{align}
    \xi(r) &= \int_0^{\infty}\frac{\dd{k}}{(2\pi)^3} 4\pi k^2 \,  \frac{ \sin(kr) }{kr} \, P(k), \\
    P_{\mathrm{cs}}^{\mathrm{1h}}(k) &= \frac{1}{\bar{n}_{\rmg}^2} \int\dd{M} \textcolor{blue}{\dv{n(M)}{M}} \expval{N_{\mathrmc}}\!(M)\, \lambda_{\mathrms}(M) \, \textcolor{blue}{u_{\mathrm{s}} (k | M)}, \label{eqn:xics_1h_exp_erfs} \\
    P_{\mathrm{ss}}^{\mathrm{1h}}(k) &= \frac{1}{\bar{n}_{\rmg}^2} \int\dd{M}\textcolor{blue}{\dv{n(M)}{M}} \expval{N_{\mathrmc}}\!(M)\, \big[\lambda_{\mathrms}(M)\big]^2 \, \big[\textcolor{blue}{u_{\mathrm{s}} (k | M)}\big]^2, \label{eqn:xiss_1h_exp_erfs}
\end{align}
where $u_{\mathrm{s}}(k|M)$ is (the Fourier transformation of) the radial distribution of satellite galaxies within a halo, and we have highlighted the emulated quantities in blue.
In this section, we assume that the distribution is given by an NFW profile with the concentration-mass relation from \citet{Diemer:2019ApJ...871..168D}.

\begin{figure*}
    \centering
    \includegraphics[width=0.5\textwidth]{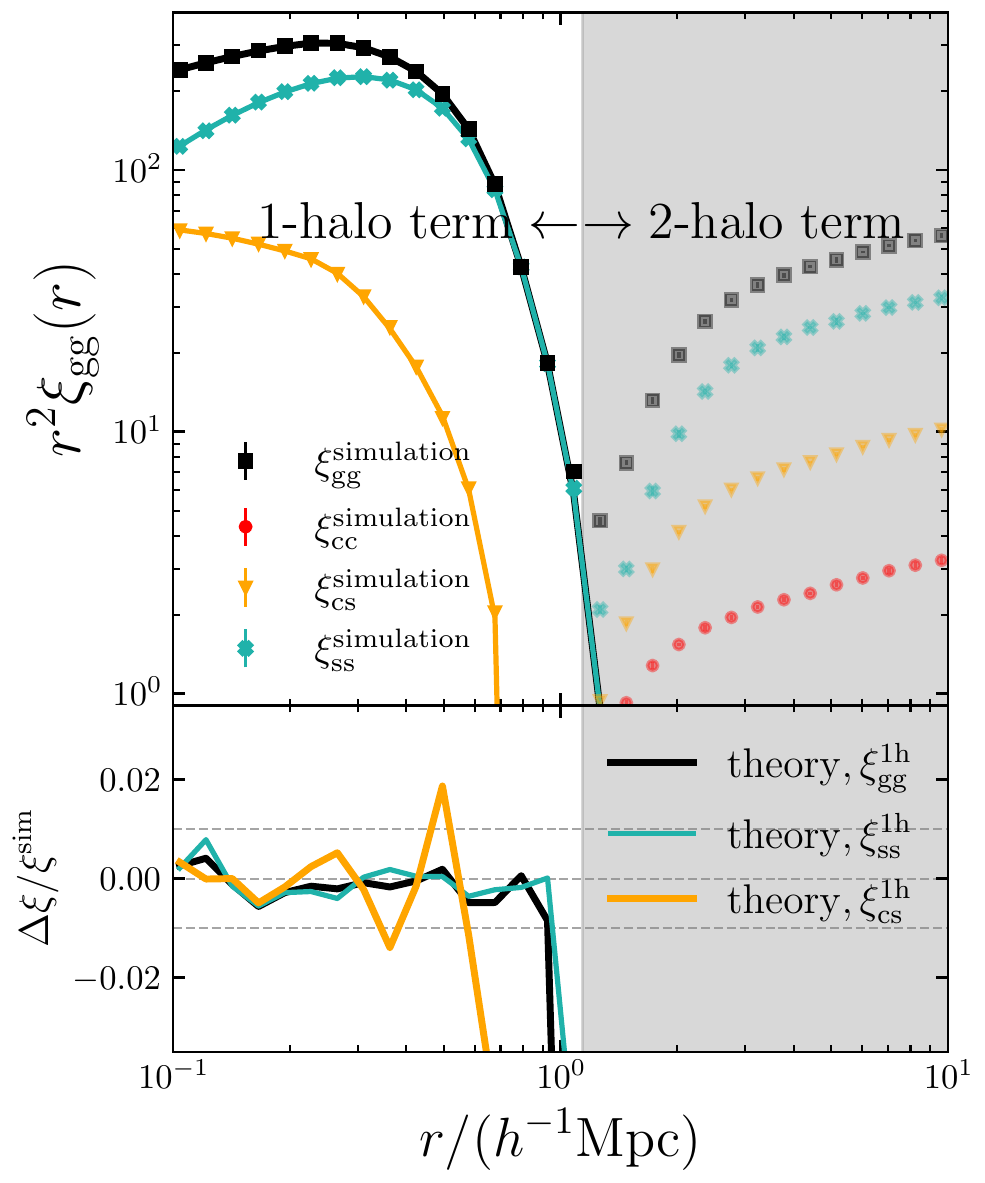}%
    \includegraphics[width=0.5\textwidth]{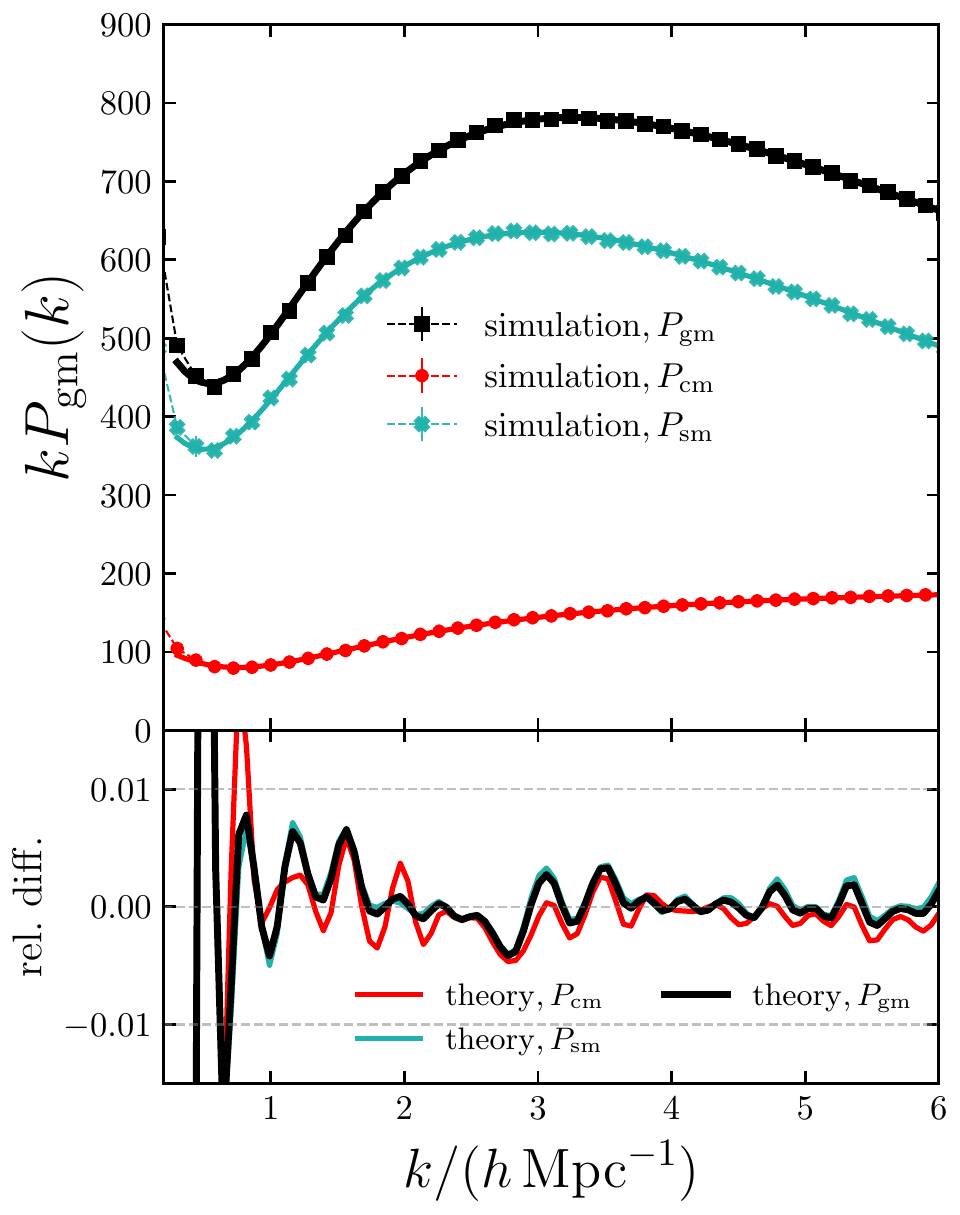}
    \caption{ 
    Emulator-based halo model predictions of galaxy clustering statistics.
    \textit{Left} panel: Galaxy two-point correlation functions from simulations (marks) and emulator-based halo model predictions (solid lines), for the fiducial \textit{Planck} cosmology of \textsc{FORGE} at $z=0$. 
    Colours represent different terms of galaxy correlations after the central/satellite split: galaxy-galaxy ({black}), central-central ({red}), central-satellite ({orange}) and satellite-satellite ({green}).
    Only the one-halo terms of theory predictions are shown.
    In the {lower} sub-panel, we show the relative difference between the 1-halo term and the full correlation function measured from simulations.
    \textit{Right} panel: Similar to the left panel but for the galaxy-matter cross power spectrum for the same cosmology and HOD prescription. 
    (\href{https://github.com/chzruan/HM_emulator_paper1_figures/blob/main/xi_gg_halo_model.py}{source code})
    }
    \label{fig:xi_gg_emuhm}
\end{figure*}

The left panel of Fig.~\ref{fig:xi_gg_emuhm} compares the model predictions and the galaxy TPCF measured from mock HOD catalogues at $z=0$.
On the scales where the 1-halo term dominates ($r \lesssim 1\,h^{-1}\mathrm{Mpc}$), the fractional difference is within $1\%$.
The colours in the plot represent different contributions: the correlations of central-central, central-satellite and satellite-satellite galaxy pairs.

\begin{figure}
    \centering
    \includegraphics[width=\columnwidth]{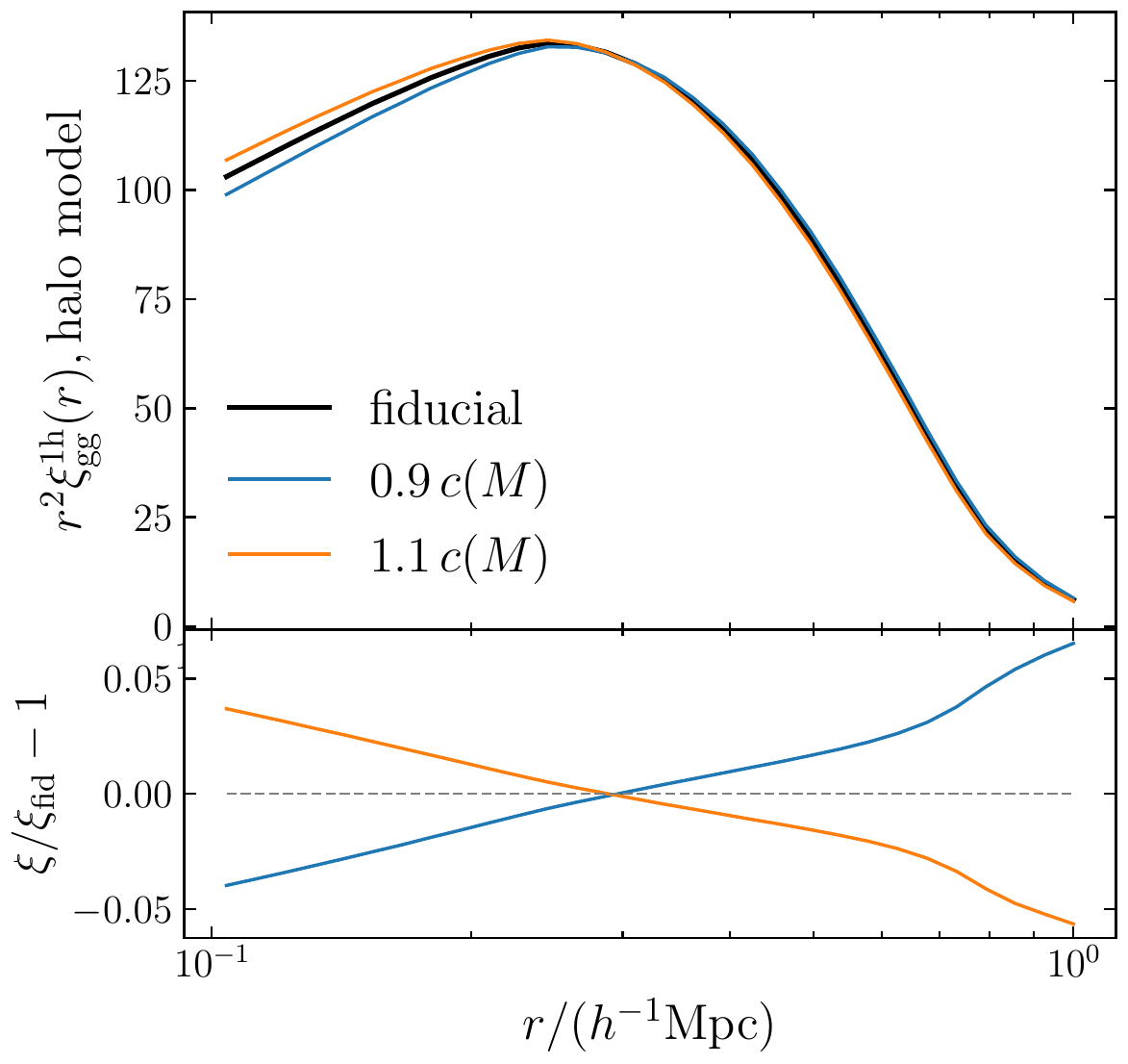}
    \caption{One-halo terms of the galaxy two-point correlation functions with the NFW profiles combined with different concentration-mass relations.
    The black line shows the fiducial case corresponding to the $c(M)$ relation measured from this work.
    The other two coloured lines present the results for increasing and decreasing the concentration by $10$ per cent.
    In the lower sub-panel, we show the relative difference with respect to the fiducial result.
(\href{https://github.com/chzruan/HM_emulator_paper1_figures/blob/main/plot_xi_gg_theory_varconc.py}{source code})
    }
    \label{fig:xi_gg_theory_varconc}
\end{figure}

As shown in Section~\ref{subsec:halo_profile_individual_nfw} and Fig.~\ref{fig:ResolutionTest_all_individual_NFW}, the halo concentrations measured from simulations are sensitive to the minimum radius in the fitting, with a relative difference of up to 10 per cent.
To test the impact on galaxy clustering, we calculate the one-halo terms of $\xi_{\rmgg}$ (Eqns.~\eqref{eqn:xics_1h_exp_erfs} and \eqref{eqn:xiss_1h_exp_erfs}), adopting the NFW profile with the concentration-mass relations measured from this work and increasing or decreasing them by $10$ per cent.
Fig.~\ref{fig:xi_gg_theory_varconc} shows that a $10$ per cent change in the concentration-mass relation will change the one-halo term by $5$ per cent.
The impact on the two-halo term and the degeneracy between the concentration amplitude and galaxy-halo connection model parameters will be left for future work.

\subsection{Galaxy-matter cross-correlation function}
\label{subsec:application_xi_gm}

In the galaxy-galaxy weak lensing observations, the excess surface mass density profile around lensing galaxies, $\Delta \Sigma_{\text{gm}}(R)$ is measured, which can be expressed in terms of the galaxy-matter cross-power spectrum as \citep[e.g.,][]{Murata:2018ApJ...854..120M,Nishimichi:2019ApJ...884...29N}
\begin{align}
    \Delta \Sigma_{\text{gm}} (R) = \bar{\rho}_{\rmm0} \int_0^{\infty} \frac{\dd{k}}{2\pi} k P_{\text{gm}}(k) J_2(kR),
\end{align}
where $J_2(x)$ is the second-order Bessel function.

In the halo model framework, we can accurately predict $P_{\text{gm}}(k)$ with the emulators providing the model ingredients.
Under the same configurations as in the last sub-section, $P_{\text{gm}}(k)$ is related to the halo properties as
\begin{align}
    P_{\text{gm}}(k) &= \frac{1}{\bar{n}_{\rmg}} \int \dd{M} {\color{blue}\dv{n(M)}{M}} \expval{N_{\mathrm{c}}}\!(M) \times \notag \\
    & \phantom{\frac{1}{\bar{n}_{\rmg}} \int \dd{M}\quad} \Big[ 1 + \lambda_{\mathrm{s}}(M) \, {\color{blue}u_{\mathrm{s}}(k | M)} \Big] {\color{blue}P_{\text{hm}}(k | M)}.
\end{align}
$P_{\text{hm}} (k | M)$ is the cross-correlation between the matter overdensity field with the halo sample in a narrow mass bin $[M, M + \dd{M}]$.
The quantity that our halo-mass cross-correlation emulators output is $P_{\text{hm}} (k | n_\rmh(>M))$, which can be converted as
\begin{align}
    P_{\text{hm}} (k | M) = -\qty[\dv{n_\rmh(M)}{M}]^{-1} \frac{\partial}{\partial M} \Big[n_\rmh(>M) P_{\text{hm}}\qty(k \big| n_\rmh(>M)) \Big].
\end{align}

We use the publicly available, open-source Python toolkit \href{https://nbodykit.readthedocs.io/en/latest/index.html}{\texttt{nbodykit}} \citep{Hand:2018AJ....156..160H.nbodykit} to measure the cross power spectra between the mock galaxy catalogues and the matter field, in linear $k$ bins from $0.3$ to $6\,h\,\mathrm{Mpc}^{-1}$ with a width of $\Delta k = 0.02\,h\,\mathrm{Mpc}^{-1}$.
These measurements are compared with the halo model predictions in the right panel of Fig.~\ref{fig:xi_gg_emuhm}.
The relative difference shown in the lower sub-panel is below 1 per cent except at low-$k$ bins, where the cosmic variance dominates the error budget.

\section{Discussions and Conclusions}
\label{sec:conclusions}


We present accurate emulators for the halo mass function, concentration-mass relation and halo-matter cross-correlation function, for $\Lambda$CDM and two representative modified gravity theories, $f(R)$ gravity and DGP, using the \textsc{FORGE} and \textsc{BRIDGE} suites of $N$-body simulations \citep{Arnold:2022MNRAS.515.4161A.FORGE_1}.
The cosmological parameter space spans three non-MG parameters, $\Omega_{\mathrm{m0}}, h, S_8$, and one MG parameter, either $\bar{f}_{R0}$ or $H_0 r_{\mathrm{c}}$, depending on which modified gravity model we are using.

We construct emulators using fully connected neural networks implemented using the open-source Python library \href{https://www.pytorchlightning.ai/}{\texttt{PyTorch Lightning}}.
We show the capabilities of neural networks under noise-free conditions by emulating the existing fitting formulae of halo properties, such as the fitting function for HMFs in \citet{Tinker:2008ApJ...688..709T.HMF}.
The emulators mimic analytical models in a 7-D parameter space with sub-percent accuracy, using only $50$ training cosmologies.

For realistic cases where the data come from $N$-body simulations and therefore have noise, the accuracy of our halo property emulators is summarised in Figs.~\ref{fig:HMF_emu}-\ref{fig:profile_NFW_emu_vs_sim_FORGE50}. 
The emulation error is less than $1\%$ for most cosmologies in both the training and the test data sets, in the halo mass range of $10^{12} \le M_{\text{200c}} / (h^{-1} M_{\odot}) \le 10^{14}$.


The primary purpose of this series of papers is to extend the modelling of galaxy clustering to non-linear scales.
We employ the halo model framework \citep{Cooray:2002PhR...372....1C} combined with an adopted galaxy-halo connection model to predict galaxy clustering and other cosmological observables, following the spirit of the \textsc{Dark Quest} project \citep{Nishimichi:2019ApJ...884...29N,Kobayashi:2020PhRvD.102f3504K,Cuesta-Lazaro:2022arXiv220805218C}.
We demonstrate that the emulators can be applied to the halo model framework combined with the HOD prescription to predict the one-halo term of the galaxy clustering signal, achieving sub-percent accuracy.
The main advantages of this emulator-based halo model approach can be summarised as follows.
\begin{itemize}
    \item Accuracy. 
    The model ingredients provided by emulators incorporate the major complicated effects in the non-linear regime of structure formation, such as non-linear halo bias and the halo exclusion effect.
    \item Versatility. 
    The halo model approach enables a joint modelling of cosmological observables, such as galaxy-galaxy and galaxy-matter correlation functions, for a single population of galaxies.
    The combination of different probes can mitigate the uncertainty of galaxy formation and evolution on cosmological parameter inference.
    \item Flexibility. 
     Instead of making an end-to-end mapping between the cosmological and HOD parameters to the final clustering statistics with the emulation process, this ``numerical'' version of the halo model allows the flexibility of combining with any specific HOD prescription for different types of galaxies, without retraining the emulators.
\end{itemize}


To perform cosmological parameter inference by confronting the emulator-based halo model prediction with galaxy survey observations, we plan to implement the following improvements and extensions in the subsequent papers of this series.
\begin{itemize}
    \item The excellent performance of the emulators is partly due to the smaller number of parameters varied in the \textsc{FORGE} and \textsc{BRIDGE} simulations compared with other emulation projects, as well as the limited halo mass range due to the relatively small box size of the simulations.
    However, as indicated by the ideal emulation tests, neural networks are capable of emulating halo properties up to the halo masses of $10^{15.5}\,h^{-1} M_{\odot}$ in a higher-dimensional parameter space with sub-percent accuracy.
    To extend the mass range of the emulators, we plan to run additional simulations with different specifications, such as box size and number of particles, to obtain halo properties robustly and at low cost.
    \item Galaxy clustering statistics are typically measured in redshift space from surveys, which involves not only the information about galaxy positions but also their peculiar velocities.
    We will build emulators for the halo peculiar velocity statistics, such as pairwise velocity moments, and combine them using a galaxy-halo connection model \citep[e.g.][]{Kobayashi:2020PhRvD.102f3504K,Cuesta-Lazaro:2022arXiv220805218C}.
    \item To resemble actual samples of galaxies, we need more realistic HOD prescriptions and check the accuracy of the emulator-based halo model.
\end{itemize}

In the future, we plan to use the neural network emulators on the upcoming data from DESI and \textit{Euclid} to constrain the cosmological and modified gravity parameters. 
This requires that the models be trained on simulations with higher resolution to meet the demand of the cutting-edge observational data with unprecedented volume and much better controlled systematic errors.

\section*{Acknowledgements}
C-ZR thanks the Research Council of Norway for their support.
C-ZR and BL are supported by the European Research Council (ERC) through a starting Grant (ERC-StG-716532 PUNCA). AE is supported at the AIfA by an Argelander Fellowship.
CMB acknowledges support from the Science Technology Facilities Council through ST/T000244/1.
BL and CMB are further supported by the UK Science and Technology Funding Council (STFC) Consolidated Grant No. ST/I00162X/1 and ST/P000541/1.
CTD is funded by the Deutsche Forschungsgemeinschaft (DFG, German Research Foundation) under Germany's Excellence Strategy -- EXC-2094 -- 390783311.

This work used the DiRAC@Durham facility managed by the Institute for Computational Cosmology on behalf of the STFC DiRAC HPC Facility (www.dirac.ac.uk). The equipment was funded by BEIS capital funding via STFC capital grants ST/K00042X/1, ST/P002293/1, ST/R002371/1 and ST/S002502/1, Durham University and STFC operations grant ST/R000832/1. DiRAC is part of the National e-Infrastructure.

\section*{Data Availability}

The data underlying this article will be shared on reasonable request to the corresponding author.
The source code and data to generate the figures in this manuscript are available at \href{https://github.com/chzruan/HM_emulator_paper1_figures}{this GitHub repository}.




\bibliographystyle{mnras}
\bibliography{emulator_cm} 







\bsp	
\label{lastpage}
\end{document}